\documentclass[a4paper,fleqn,usenatbib]{mnras}

\usepackage{amsmath}	% Advanced maths commands
\usepackage{amssymb}	% Extra maths symbols
\usepackage{newtxtt,newtxmath}
\usepackage[T1]{fontenc}
\DeclareRobustCommand{\VAN}[3]{#2}
\let\VANthebibliography\thebibliography
\def\thebibliography{\DeclareRobustCommand{\VAN}[3]{##3}\VANthebibliography}

\usepackage{graphicx}	% Including figure files
\usepackage{epstopdf}
\usepackage{subcaption}
\usepackage{multicol}
\usepackage{multirow}
\usepackage{float}
\usepackage[usenames,dvipsnames,svgnames,table]{xcolor}
\usepackage{enumitem}
\usepackage{bm}
\usepackage{tablefootnote}
\usepackage{orcidlink}

\usepackage[noabbrev]{cleveref}

\crefname{equation}{}{}
\crefname{figure}{}{}

\newcommand{\wo}{$W_0$}
\newcommand{\wom}{$W_0(M,z)$}

\newcommand{\lya}{Ly$\alpha$}
\newcommand{\ha}{H$\alpha$}

\newcommand{\oiii}{[O{\sc iii}]}
\newcommand{\oii}{[O{\sc ii}]}
\newcommand{\nii}{[N{\sc ii}]}
\newcommand{\ewr}{EW$_{\mathrm{0}}$}
\newcommand{\msol}{M$_\odot$}

\newcommand{\aha}{$A_\textrm{\ha}$}
\newcommand{\cgsline}{erg s$^{-1}$ cm$^{-2}$}
\newcommand{\ionxi}{$\xi_\textrm{ion}$}

\setlist[enumerate]{leftmargin=*} 

\def\arraystretch{1}% 
\title[Evolution of \ha~EWs]{Evolution of \ha~Equivalent Widths from $z \sim 0.4 - 2.2$: implications for star formation and legacy  surveys with \textit{Roman} and \textit{Euclid}}
\author[Khostovan et al.]{A.~A.~Khostovan$^{1,2}$\thanks{NASA Postdoctoral Program Fellow}\thanks{E-mail:
		akhostov@gmail.com}\orcidlink{0000-0002-0101-336X},  
		S.~Malhotra$^{2}$\orcidlink{0000-0002-9226-5350}, 
		J.~E.~Rhoads$^{2}$\orcidlink{0000-0002-1501-454X}, 
		D.~Sobral$^{3,4}$\orcidlink{0000-0001-8823-4845}, 		
		S.~Harish$^{1}$\orcidlink{0000-0003-0129-2079},
		V.~Tilvi$^{5}$\orcidlink{0000-0001-8514-7105},\newauthor
		A.~Coughlin$^{6}$\orcidlink{0000-0002-8147-0834},
		S.~Rezaee$^{7}$\orcidlink{0000-0002-6289-9918}\\
	$^{1}$School of Physics and Astronomy, Rochester Institute of Technology, 84 Lomb Memorial Drive, Rochester, NY 14623, USA\\	
	$^{2}$Astrophysics Division, NASA Goddard Space Flight Center, Greenbelt, MD 20771, USA\\
	$^{3}$Department of Physics, Lancaster University, Lancaster LA1 4YB, UK\\
	$^{4}$BNP Paribas Corporate \& Institutional Banking, Torre Ocidente Rua Galileu Galilei, 1500-392 Lisbon, Portugal\\
	$^{5}$Directorate of Higher Education, Porvorim Goa 403521 India\\	
	$^{6}$Chandler-Gilbert Community College, 2626 East Pecos Road, Chandler, AZ 85225-2499, USA\\
	$^{7}$Department of Physics and Astronomy, University of California, Riverside, CA 92521, USA
}
\date{}

\pubyear{2024}
\begin{document}

\label{firstpage}
\pagerange{\pageref{firstpage}--\pageref{lastpage}}
\maketitle

% Abstract of the Paper
\begin{abstract} 
Past studies have investigated the evolution in specific star formation rate (sSFR) and its observational proxy (\ha~equivalent width; EW) up to $z \sim 6$; however, such measurements are prone to selection biases that may overestimate the typical sSFR/EW at a given redshift. We investigate the `intrinsic' (corrected for selection and observational effects) \ha~EW distributions of $z \sim 0.4 - 2.2$ narrowband-selected \ha~samples from HiZELS and DAWN using a forward modeling approach where we assume an `intrinsic' exponential EW distribution, apply selection and filter effects, and compare with observed \ha~EW distributions. We find an `intrinsic' EW -- stellar mass anti-correlation, $M^\gamma$, with steepening slopes $\gamma = -0.18\pm0.03$ to $-0.24^{+0.06}_{-0.08}$ at $z \sim 0.4$ and $z\sim 2.2$, respectively. Typical rest-frame \ha~EW~increases as $(1+z)^{1.78^{+0.22}_{-0.23}}$ for a $10^{10}$ \msol~emitter from $15^{+2.4}_{-2.3}$\AA~($z \sim 0.4$) to $67.7^{+10.4}_{-10.0}$\AA~($z \sim 2.2$). The \ha~EW redshift evolution becomes steeper with decreasing stellar mass highlighting the high EW nature of low-mass systems at high-$z$. We model this redshift evolving EW--stellar mass anti-correlation, \wom, and find it produces \ha~luminosity functions and SFR functions strongly consistent with observations validating the model. This allows us to use our best-fit \wom~model to investigate the relative contribution of \ha~emitters towards cosmic star formation at a given epoch. We find \ewr~$>200$\AA~emitters (potentially bursty systems) contribute significantly to overall cosmic star formation activity at $z \sim 1.5 - 2$ making up $\sim 40$\% of total cosmic star formation consistent with sSFR $> 10^{-8.5}$ yr$^{-1}$ (mass doubling times of $\lesssim 300$ Myr) which we find makes up $\sim 45 - 55$\% of cosmic star formation at $z \sim 2$. Overall, this shows the importance of high EW systems in the high-$z$ Universe. Our best-fit \wom~model also reproduces the cosmic sSFR evolution found in simulations and observations (coupled with selection limits typically found in narrowband surveys), such that tension between the two can simply arise from selection effects in observations. Lastly, we forecast {\it Roman} and {\it Euclid} grism surveys using our best-fit \wom~model and include the effects of limiting resolution and observational efficiency. We predict number counts of $\sim 24000$ and $\sim 30000$ $0.5 < z < 1.9$ \ha~emitters per deg$^{-2}$, respectively, down to $>5\times10^{-17}$ \cgsline~including $10^{7.2 - 8}$ \msol~galaxies at $z > 1$ with \ewr~$>1000$\AA. Both {\it Roman} and {\it Euclid} will enable us to observe with unprecedented detail some of the most bursty/high EW, low-mass star-forming galaxies near cosmic noon.
 \end{abstract}


\section{Introduction}

Constraining the star-formation histories of galaxies is crucial in understanding their stellar mass buildup over cosmic time and how they become the large, present-day structures in the local Universe. Star-formation history is generally smooth with galaxies residing on a linear correlation between their star-formation rates (SFRs) and stellar mass (e.g., `Main Sequence'; \citealt{Daddi2007,Noeske2007,Speagle2014}). However, a unique subset of galaxies reside above this correlation with elevated SFR at a given stellar mass and are known as starbursts given their enhanced burst of star formation activity. Understanding how bursty star-formation activity occurs, how dominant or minor this mode of star formation was at various epochs, and what its overall contribution to total cosmic star-formation activity was is important in understanding the evolutionary path of galaxies into the present structures we see in the local Universe.

One commonly used tracer to identify sources that may have recent bursty star-formation activity is the \ha/UV ratio where \ha~traces instantaneous star-formation ($\sim 3 - 10$ Myr; \citealt{Kennicutt2012}) via bright, massive, short-lived $O$ and $B$ type stars and UV continuum light tracing longer periods of SF activity ($\sim 100$ -- $200$ Myr; \citealt{Kennicutt2012}). An elevated \ha~emission in respect to UV continuum flux would, in principle, signify a star-forming galaxy undergoing some recent, bursty star-formation activity in the past $\sim 10$ Myr. This approach has been used to study burstiness\footnotemark{} of galaxies in the local (e.g., \citealt{Glazebrook1999,Weisz2012, Emami2019}), $z < 2$  (e.g., \citealt{Guo2016,Atek2022,Mehta2022,Patel2023}), and $z > 2$ Universe  (e.g., \citealt{Faisst2019}). However an elevated \ha/UV ratio can also result from a varying IMF (e.g., \citealt{Boselli2009,Meurer2009,Pflamm2009}), underestimation (overestimation) of UV (\ha) dust correction, and variations in the nebular-stellar dust reddening (e.g., \citealt{Lee2009,Reddy2015,Theios2019}). \ha~emission can also be affected by the escape fraction of ionizing photons (e.g., \citealt{Matthee2017b}). Furthermore, binary stellar populations may increase the lifetime of massive $O$ and $B$ type stars raising the SF timescale of \ha~emission to be comparable with UV  (e.g., \citealt{Eldridge2017}).

While \ha/UV has been used for estimating recent star formation and potential burstiness, a recent study by \citet{Rezaee2022} used a sample of 310 $z \sim 2$ star-forming galaxies with Keck/MOSFIRE and LRIS coverage and high-resolution, rest-frame UV {\it HST} and found no correlation between \ha/UV and SFR surface density ($\Sigma_\textrm{SFR}$). \citet{Rezaee2022} also finds a lack of elevated C{\sc iv} and Si{\sc iv} emission and P Cygni features with increasing \ha/UV ratio which would have been indicative of the presence of massive stars and stellar driven winds and concludes that \ha/UV ratio may not be a reliable tracer of bursty star-formation activity. We therefore need a more reliable indicator of burstiness, especially for high-$z$ star-forming galaxies.

\footnotetext{The term `burstiness' is used commonly within the literature but with differences in regards to its meaning. In the context of this work, we refer to `burstiness' and `bursty star formation' as a galaxy that has undergone a rapid and intense period of star-formation activity relative to its past star formation. This is also traced by both the instantaneous sSFR and emisison line EW which is defined as the ratio of recent to past star-formation activity. Other definitions of `burstiness' include the variations of SFR factoring in both the rise and fall of star formation activity encompassing its full SFH. This should not be confused with our definition as mentioned above.}

\ha~equivalent width (EW) is an observational alternative to investigate the recent star-forming histories of galaxies and has several advantages. First, \ha~EW is by definition an observational proxy of specific star formation rate (sSFR) and, therefore, traces the doubling mass timescale from recent star-formation activity (e.g., \citealt{Fumagalli2012}). Second, it is even more sensitive to recent star-formation activity as it is the ratio between \ha~emission and rest-frame $R$ continuum (traces even older stellar populations; timescales $> 100$Myr; \citealt{Wang2020}). Lastly, \ha~EW is, essentially, independent of dust correction if one assumes a nebular-stellar reddening of unity ($f = E(B -V)_\textrm{stellar}/E(B-V)_\textrm{nebular}$, $f \sim 1$) although assuming different $f$ values will result in systematically different \ha~\ewr~with factor $10^{0.4 k(\textrm{6563\AA}) E(B-V)_\textrm{stellar} [ 1/f  - 1]}$, where $k(6563\AA)$ is the reddening curve at 6563\AA. Overall, \ha~equivalent width has several key advantages over \ha/UV as a tracer of bursty star-formation activity.

The distribution of \ha~EW in a sample of galaxies would qualitatively show the range of recent star-formation history from normal to bursty star-forming galaxies. Past studies of \ha~EW distributions find an anti-correlation with stellar mass with a redshift evolution where low-mass galaxies tend to have higher EWs and the typical EWs  increase with redshift as $(1+z)^{1.8}$ up to $z \sim 2$ and $(1+z)^{1.3}$ from $z \sim 2 - 6$ (e.g., \citealt{Fumagalli2012,Sobral2014,Faisst2016,Marmol2016,Rasappu2016,Reddy2018}). This suggests for wider \ha~EW distributions and increasing number of bursty star-forming galaxies with increasing redshift and at lower stellar masses. This implies that bursty star-forming galaxies may be an important contributor to overall star-formation activity in the high-$z$ Universe. For example, \citet{Atek2014} used a sample of 1000 $0.3 < z < 2$ \ha~emitters with \textit{HST} grism spectroscopy and found that rest-frame EW$>100$\AA~sources contributed $\sim 34$\% to the overall $z \sim 1  - 2$ cosmic star-formation rate density. Recent {\it JWST} observations are also finding a ubiquitous population of high \ha~EW systems \citealt{Sun2022,Endsley2023,Matthee2023,Rinaldi2023}). Interpreting \ha~EW as a tracer of burstiness, these results may suggest rapid intense periods of star formation activity may be a common mode of star formation activity in the high-$z$ Universe. \citet{Tran2020} suggest that a bursty phase of star-formation in the high-$z$ Universe would result in high EW systems and such a feature would decrease with decreasing redshift as the stellar continuum increases with the overall stellar mass build-up.

However, past studies of typical \ha~EWs, its anti-correlation with stellar mass, and its redshift evolution may be overestimated simply due to selection biases not being taken into account. We demonstrated this in \citet{Khostovan2021} where we used the $z = 0.47$ \ha~narrowband-selected LAGER sample \citep{Khostovan2020} with a forward modeling approach also used in this study (see \S\ref{sec:modeling}) to investigate how selection effects could affect the underlying measurements derived from \ha~EW distributions. In that study, we found an intrinsic (selection-corrected) EW$_0(\textrm{\ha}) \propto M^{-0.16\pm0.03}$ with $>5\sigma$ significance from a null correlation such that an intrinsic anti-correlation between \ha~EW and stellar mass exists. This intrinsic correlation can also recover the \ha~luminosity function coupled with a stellar mass function. However, not correcting for selection effects results with a steeper correlation of EW$_0(\textrm{\ha}) \propto M^{-0.25\pm0.04}$ in strong agreement with past studies (e.g., \citealt{Sobral2014}) and does not recover the \ha~luminosity function. 

\begin{table*}
	\centering
	\caption{List of narrowband surveys used in this study along with the corresponding filter parameters, sample redshifts, and sample size. Area is defined as the effective sky coverage, which removes masked regions. The effective comoving volume also incorporates the removal of masked regions.}
	{\renewcommand{\arraystretch}{1.3}
	\begin{tabular*}{\textwidth}{@{\extracolsep{\fill}} l l c c c c c c c c}
		\hline
		Survey 	& Reference & Instrument & Filter & $\lambda_c$ & FWHM & $z$ & $N_{gal}$ & Area & Volume \\
		&				&		 &                     &  ($\micron$) & (\AA) & & &  (deg$^2$) & (10$^4$ Mpc$^{-3}$)  \\ 
		\hline						
		HiZELS & \citet{Sobral2013} & Subaru/SuprimeCam & NB921 & 0.9196 & 132 & 0.40 & 1123 & 1.86 & 8.8\\
		&						   &  UKIRT/WFCAM & NBJ      & 1.211 & 150 & 0.84 & 637 & 1.30 & 19.1\\
		&						   &  UKIRT/WFCAM & NBH    & 1.617 & 211 & 1.47 & 515 & 2.16 & 73.6\\
		&						   &  UKIRT/WFCAM & NBK	 & 2.121 & 210 & 2.23 & 772 & 2.18 & 77.2 \\
		%CF-HiZELS & \citet{Sobral2015} & 0.81 & 2834 & 9.2 & 86 & 3 & 25\\
		DAWN  & \citet{Coughlin2018}, & Mayall/NEWFIRM & NB1066 & 1.066 & 35 & 0.62 & 241 & 1.5 & 3.5 \\
		& \citet{Harish2020} &  &  &  &  &  &  & & \\	
		\hline
	\end{tabular*}
	}
\label{table:surveys}
\end{table*}

In order to properly gauge the \ha~EW redshift evolution and anti-correlation with stellar mass, we need to carefully take into account and correct for selection effects. In this study, we extend our work from \citet{Khostovan2021} by applying our intrinsic \ha~EW modeling via our forward modeling approach on the narrowband-selected High-$z$ Emission Line Survey (HiZELS; \citealt{Geach2008,Sobral2013}) and the Deep And Wide Narrowband survey (DAWN; \citealt{Coughlin2018,Harish2020}). Both surveys are currently the largest near-IR narrowband surveys available and have several key advantages for the purpose of this work. The samples are selected based on emission line contribution within the narrowband filter and have measurement for emission line flux, continuum flux density, and equivalent width. The selection functions are also very simple to model, typically a combination of line flux and EW-limits as demonstrated in \citet{Khostovan2021}. The emission line-selection also limits redshift uncertainties to the narrowband FWHM and eliminates any internal redshift evolution in the sample (e.g., a sample that has a wide redshift bin width). Lastly, the combination of HiZELS and DAWN covers from $z \sim 0.4$ to $\sim 2.2$ allowing for a comprehensive overview of the changing star-formation histories of star-forming galaxies from high-$z$ to low-$z$. Specifically, we would address the role of bursty star-forming galaxies at different cosmic times in regards to cosmic star-formation activity. Lastly, we will use our modeling approach to also present predictions for the range of \ha~EW planned legacy surveys with \textit{Roman} and \textit{Euclid} can observe given survey conditions (e.g., flux limits, grism resolution, survey area, etc.) 

The paper is structured as follows: the samples are defined in \S\ref{sec:samples}. We then provide an overview of our forward modeling approach in \S\ref{sec:modeling} and refer the reader to \citet{Khostovan2021} for more details. \S\ref{sec:results} highlights our results in the study where we present measurements of the intrinsic, selection-corrected \ha~EW -- stellar mass anti-correlation and its redshift evolution, show how it can reproduce both the \ha~luminosity function and SFR function, how different types of star-forming galaxies contribute to overall cosmic star-formation activity at different cosmic times, and how selection effects can augment measurements of the cosmic \ha~EW and sSFR evolutions causing disagreement with simulations. In \S\ref{sec:forecast} we predict the number counts and range of \ha~EW that can be observed with  \textit{Roman} and \textit{Euclid}. In \S\ref{sec:discussion}, we present discussions on the implications our work has on survey planning, how starburst galaxies may be important sources in the high-$z$ Universe, what our results imply regarding the potential of high \ha~EW emitters to be potential sources of reionizing the Universe at $z > 6$, and how our constraints on the \ha~EW distribution can help in assessing low-$z$ contaminants/interlopers in high-$z$ \lya~emitter-selected samples. Lastly, we present our main conclusions in \S\ref{sec:conclusion}.

Throughout this paper, unless otherwise stated, we assume $\Lambda$CDM cosmology ($H_0 = 70$ km s$^{-1}$ Mpc$^{-1}$, $\Omega_m = 0.3$, $\Omega_\Lambda = 0.7$) and a \citet{Chabrier2003} initial mass function (IMF). Magnitudes assume the AB magnitude system.

\section{Samples}
\label{sec:samples}

In this work, we use samples derived from two of the largest, ground-based near-IR narrowband surveys described below. Survey parameters and sample sizes are listed in Table \ref{table:surveys}.

\begin{figure*}
	\centering
	\includegraphics[width=\textwidth]{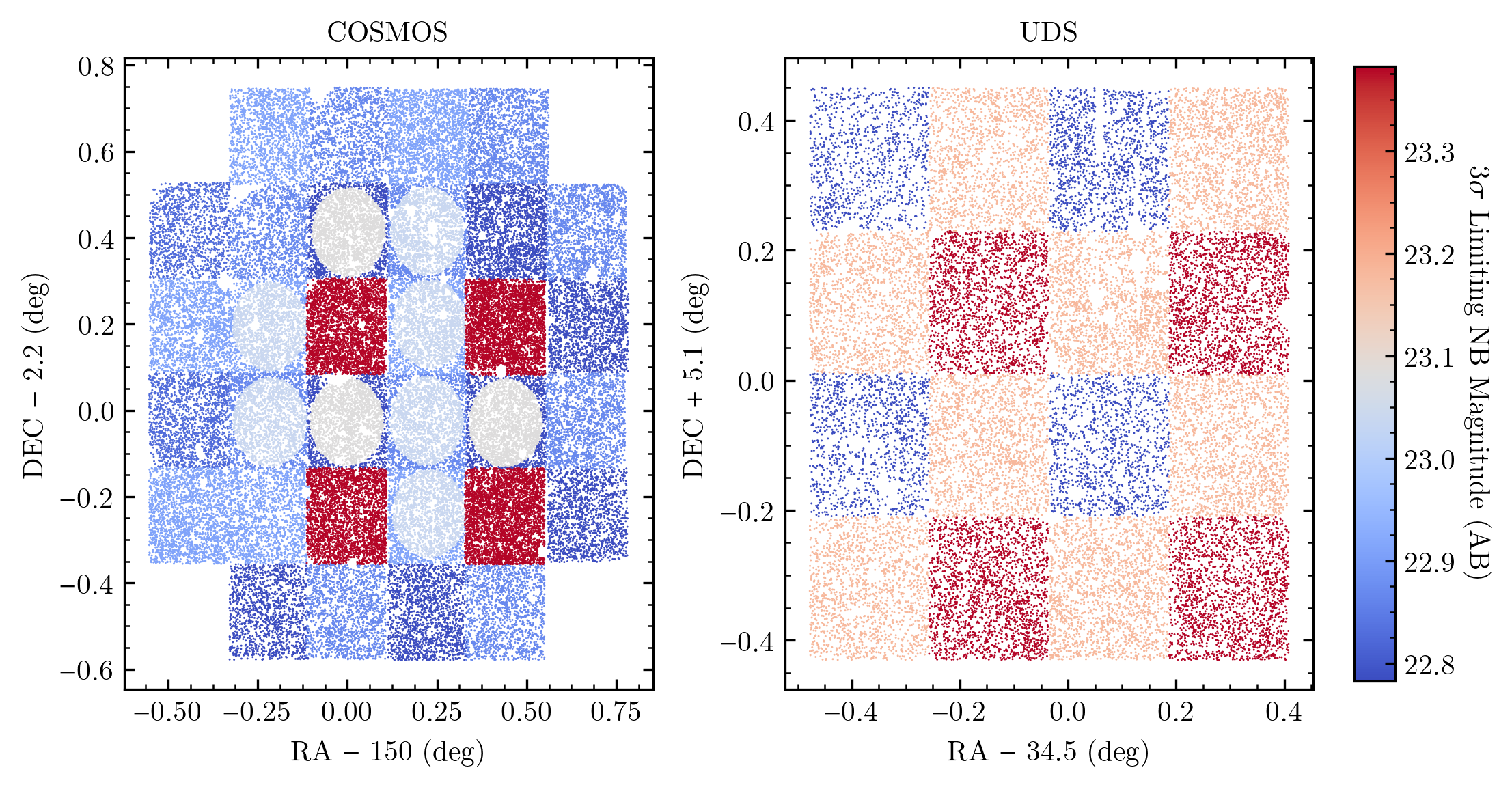}
	\caption{Source distribution of all HiZELS NBH detected sources highlighting the inhomogeneous depths in the samples used in this study. Both fields have 3$\sigma$ limiting NB magnitudes varying by $\sim 0.5$ mag. Each pointing also can have different effective volumes given the presence of masked regions that remove the effects of artifacts and poor photometry. Varying depth and effective volume is taken into account in our forward modeling by generating \ha~emitters down to a certain depth and volume for each pointing. This is crucial in order to form a mock \ha~sample that best represents observations otherwise we could overestimate/underestimate the number of sources in the field.}
	\label{fig:HiZELS_NBH}
\end{figure*}

\subsection{HiZELS}
\label{sec:HiZELS}

The High-$z$ Emission Line Galaxy Survey (HiZELS; \citealt{Geach2008,Sobral2013}) is a large 2 deg$^2$ narrowband survey covering COSMOS and UDS/SXDS using four narrowband filters on Subaru/SuprimeCam and UKIRT/WFCAM. The four narrowband filters were used to select \ha~emitters at $z \sim 0.4 - 2.2$ \citep{Geach2008,Sobral2009,Sobral2012,Sobral2013}, as well as \oiii~ and \oii~emitters at $z \sim 0.84 - 3.2$ and $z \sim 1.5 - 4.7$, respectively \citep{Khostovan2015}, in 4 discrete redshift slices of $\Delta z \sim 0.01 - 0.03$ in width. We use the \ha~samples as selected by \citet{Sobral2013} for this work and refer the reader to the specific study for details regarding the survey design and sample selection.

Briefly, the sample selection used in HiZELS is uniform throughout all four narrowband samples. The selection function is threefold: 1) a $3\sigma$ narrowband magnitude (line flux) limit, 2) a rest-frame EW cut of $25$\AA, and 3) a color significance cut\footnotemark of $\Sigma > 3$. The narrowband magnitude limit is imposed to remove sources with poor S/N. The equivalent width cut is set to remove any sources that could potentially mimic an emission-line within the narrowband filter profile (e.g., continuum features). The last selection limit is the color significance cut which removes sources where the nebular color excess (broadband -- narrowband) is not significant in respect to photometric errors. This essentially removes sources that may have emission lines present within the narrowband filter, but are not statistically significant above a threshold (e.g., $\Sigma > 3$ can be thought of as 3$\sigma $ significance). 

\footnotetext{The color significance, $\Sigma$, is the ratio of the broadband minus the narrowband flux densities (proxy for emission line flux) versus the associated photometric errors in both bands added in quadrature. Essentially it is a signal-to-noise measurement typically used in narrowband surveys where a $\Sigma > 3$ cut would translate as a S/N $> 3$ limit in the emission line flux uncertainies. This ensures a clean sample and the removal of sources mimicing emission line galaxies simply due to photometric uncertainties in the narrowband and/or broadband.}

Selection is done on a pointing-to-pointing basis to take into account varying depths per chip (see Figure \ref{fig:HiZELS_NBH} for HiZELS/NBH). We refer the reader to Table 2 of \citet{Sobral2013} for further details on exposure times and 3$\sigma$ limiting magnitudes for each pointing. The NB921 imaging was done using Subaru/SuprimeCam and is homogeneous in depth for the 4 pointings covering COSMOS, while the central pointing in UDS is 0.6 mag fainter compared to the other 4 pointings. NB921 UDS is also $\sim 2$ mag deeper than the corresponding COSMOS imaging. The other 3 HiZELS narrowband samples (NBJ, NBH, and NBK) were observed using UKIRT/WFCAM which consists of 4 chips per pointing. Figure \ref{fig:HiZELS_NBH} shows the 3$\sigma$ NBH depth inhomogeneity which can vary by $\sim 0.5$ mag. Each pointing/chip can also have varying effective comoving volumes in respect to other pointings/chips due to masked regions. All of these factors need to be taken into account when we generate mock samples that will be used in assessing the intrinsic EW distributions for each sample.

\subsection{DAWN}
The Deep And Wide Narrowband (DAWN) survey is a near-IR narrowband imaging survey of COSMOS, UDS, and EGS using a single, customized NB1066 filter ($\lambda_c = 10660$\AA, FWHM$ = 35$\AA) on NEWFIRM installed on the 4 m KPNO/Mayall telescope. \citet{Coughlin2018} was first to investigate the $z = 0.62$ \ha~properties using a single, deep $0.22$ deg$^2$ pointing in COSMOS selecting 116 \ha~emitters reaching a $5\sigma$ \ha~line flux limit of $\sim 1 \times 10^{-17}$ erg s$^{-1}$ cm$^{-2}$. \citet{Harish2020} extended this work by incorporating 8 shallow, flanking fields around the deep COSMOS pointing increasing the total areal coverage to $\sim 1.5$ deg$^2$. The 5$\sigma$ depth of the deep pointing is 23.6 mag while the surrounding flanking fields have 5$\sigma$ depths varying between 20.1 and 22.1 mag corresponding to \ha~line flux limits of $5 - 30 \times 10^{-17}$ erg s$^{-1}$ cm$^{-2}$. We refer the reader to Table 1 and Figure 1 of \citet{Harish2020} for more details regarding the survey design and pointing properties. A total of 241 $z = 0.62$ \ha~emitters were selected by \citet{Harish2020} using the following selection: rest-frame EW$_0 > 11$\AA~(observer-frame EW$ > 18$ \AA), $5\sigma$ limiting magnitude cuts, and $\Sigma > 3$.

\section{Forward Modeling of EW Distributions}
\label{sec:modeling}
In this section, we outline the methodology used in measuring intrinsic \ha~EW distributions following the approach outlined in \citet{Khostovan2021} and refer the reader to that study for further details.

\subsection{Brief Overview of Methodology}
\label{sec:approach}
We use the approach outlined in \citet{Khostovan2021} which was used for measuring the intrinsic \ha~EW distribution of $z = 0.47$ \ha~emitters from the LAGER survey \citep{Khostovan2020}. The approach models \ha~emitters by randomly sampling from an assumed intrinsic EW distribution and stellar mass function to generate an intrinsic \ha~sample with rest-frame EW, stellar mass, and \ha~luminosity, where the latter is inferred from the EW and stellar mass (assuming a mass-to-continuum luminosity model). \citet{Khostovan2021} found that an exponential rest-frame EW profile best represents observations of \ha~EW distributions and is defined as:

\begin{eqnarray}
p({\rm EW}|W_0) = \frac{1}{W_0} e^{-({\rm EW}/W_0)}
\label{eqn:exp_ew}
\end{eqnarray}
where \wo~is the characteristic EW and describes both the average EW for a sample (EW$ = [0 - \infty)$\AA) and the width of the exponential distribution. Higher \wo~corresponds to wider EW distributions and includes higher rest-frame EWs. We also assume an exponential distribution profile in this study. 

We generate a mock sample by first assigning an intrinsic, rest-frame EW and stellar mass, where the former is randomly assigned by sampling the intrinsic, exponential rest-frame EW distribution, as defined in Equation \ref{eqn:exp_ew} with an assumed \wo, and the latter by randomly sampling from an assumed stellar mass function (see \S \ref{sec:smf}). Rest-frame $R$-band continuum luminosity, $L_R$, is determined by combining the assigned stellar mass with our redshift-dependent $M/L_R$ calibration and assigning $L_R$ with a $0.1 - 0.2$ dex scatter (depending on the redshift) to take into account variations in the IMF and star-formation histories at a given stellar mass (see \S \ref{sec:ML_ratio}). The combination of $L_R$ and rest-frame EW determines the line flux/luminosity and the contribution of \nii~is determined by using an empirical, observationally-driven calibration (see \S\ref{sec:nii_corr}). We then incorporate the narrowband filter transmission curve effects by convolving the modeled line flux and continuum flux densities with the narrowband and broadband filters to get the observed magnitudes (see \S\ref{sec:filter_effect}). The mock photometry will be used as part of the selection process to measure line flux, rest-frame EW, and continuum flux densities using the same techniques applied in typical narrowband surveys (e.g., assuming top-hat approximations, see \S\ref{sec:filter_effect}).

With each mock sample, we convolve the selection functions corresponding to the observed samples, as described in \S\ref{sec:samples}. These selection functions include an observed \ha+\nii~line flux limit cut (narrowband magnitude limit), observer-frame \ha+\nii~EW cut, and a color significance cut. This results in mock samples having incompleteness due to selection that best represent the observed samples which will be used in this study to constrain intrinsic EW distributions. The observer-frame \ha+\nii~EW cut places a selection limit that primarily affects bright continuum (high-mass) galaxies. The \ha+\nii~line flux and color significance limits dominate at faint continuum (low-mass) galaxies and translate to an increasing EW cut with decreasing stellar mass/continuum luminosity (inversely proportional to EW). This can arbitrarily form an EW -- stellar mass anti-correlation due to selection effects alone, although we show in \citet{Khostovan2021} that a shallower, but statistically significant ($> 5\sigma$) anti-correlation is present when said effects are taken into account.

The mock samples described above are used to compare with the rest-frame EW distributions from the observed samples described in \S\ref{sec:samples}. This is done by first subdividing both the observed and mock samples in bins of rest-frame $R$-band luminosity, $L_R$. We then compare the rest-frame EW distributions of the mocks with the distributions from the observed samples per each $L_R$ bin using a $\chi^2$ minimization approach where \wo~is augmented to reach a minimum $\chi^2$ and the observed errors are Poisson. The best-fit \wo~is essentially the $e$-scaling factor that defines the shape/width of the intrinsic EW distribution before incompleteness from selection effects. The intrinsic \wo~also represents the mean EW of the sample at a given $L_R$.

In the following sections, we describe in detail the varying components and assumptions made in our approach.

\subsection{Stellar Mass Function Assumption}
\label{sec:smf}

One of the main starting points of our approach is populating our mock samples by randomly drawing \ha~emitters from an assumed stellar mass function (SMF). These are then converted to rest-frame $R$-band luminosity using an empirical model described in \S\ref{sec:ML_ratio}, which is an observable in narrowband surveys. It is crucial that the assumed SMF used in generating the mock samples be consistent with the observations. 

We follow the same approach as in \citet{Khostovan2021} by assuming the $z \sim 0.4 - 2.2$ \ha~HiZELS SMFs of \citet{Sobral2014}. These SMFs are based directly on \ha-selected emitters and represent the underlying population of star-forming galaxies. The observed samples we use in this study are also drawn primarily from HiZELS (see \S\ref{sec:HiZELS}) and follow the same selection function and thin redshift slices.

\subsection{\nii~contamination correction}
\label{sec:nii_corr}

Narrowband filters are powerful observational tools to capture \ha~emission lines and other strong nebular emission lines in blind surveys. However, the resolving power is not high enough to differentiate between \ha~and the neighboring \nii6541,6583\AA~emission lines. This is also true for blind slitless grism surveys (e.g., \citealt{Pirzkal2013,Momcheva2016,Pirzkal2017}) and for strong EW systems selected using nebular broadband excess (e.g., \citealt{Smit2014,Faisst2016}). Although assuming a constant \nii/\ha~ratio to correct for \nii~contamination would work within first order, studies of the BPT diagnostic clearly show \nii/\ha~ratios varying by a dex for star-forming galaxies (e.g., \citealt{Steidel2014}). Starburst galaxies are reported to have low \nii/\ha~ratios indicative of low gas-phase metallicities in comparison to `normal' star-forming galaxies (e.g., \citealt{Maseda2014}) and the typical \nii/\ha~ratios of galaxies decreases with increasing redshift (e.g., \citealt{Faisst2018}). 

In this study, we provide results based on both \ha+\nii~and \ha~EWs, where the latter takes the \nii~contamination into account. We adopt the \citet{Sobral2015} EW calibration which is a simplification of the SDSS-based calibration \citep{Villar2008,Sobral2009,Sobral2012}. The latter is a 5th degree polynomial fit to the \nii/\ha~line ratios of SDSS-selected \ha~emitters as dependent on rest-frame \ha+\nii~EW and is defined as:
\begin{eqnarray}
\log_{10} {\textrm\nii}/{\textrm\ha} = & -0.924 + 4.802x - 8.892x^2 \label{eqn:EW_SDSS} \\
& +6.701x^3-2.27x^4+0.279x^5 \nonumber
\end{eqnarray}
where $x = \log_{10} $EW$_0({\textrm\ha}+{\textrm\nii})$. This calibration and comparable polynomial fit calibrations have been used extensively in \ha~narrowband surveys (e.g., \citealt{Sobral2013,Matthee2017,Khostovan2020}). 

A simplification of this calibration was presented by \citet{Sobral2015} which used median stacked \nii/\ha~ratios of $z = 0.84$ \ha~emitters with Keck/MOSFIRE and Subaru/FMOS spectra in rest-frame \ha+\nii~EW bins. They found the median line ratios were consistent with the local SDSS calibration with minimally elevated \nii/\ha~ratios at EW(\ha+\nii)$\sim 25$\AA, although still consistent within $1\sigma$. The simplified, linear calibration is defined as:
\begin{eqnarray}
f({\textrm\nii}/{\textrm\ha}) = -0.296 \log_{10} \textrm{EW}_0({\textrm\ha}+{\textrm\nii}) +0.8
\label{eqn:n2_corr}
\end{eqnarray}
for rest-frame EW(\ha+\nii) between $\sim 15$\AA~and $\sim 400$\AA~where $f({\textrm\nii}/{\textrm\ha})$ is the \nii/\ha~flux ratio. We take into account this limited range in our mock samples by setting the \nii~contribution being constant at the bounds (e.g., EW$ < 15$\AA~mock emitters have \nii/\ha~ratios fixed at 2.8 consistent with EW$ = 15$\AA~using Equation \ref{eqn:n2_corr}). The key advantage of using Equation \ref{eqn:n2_corr} is the reduced number of free parameters in comparison to Equation \ref{eqn:EW_SDSS} and the focus on observables being the \ha+\nii~EW.

We implement the \citet{Sobral2015} calibration in our mock samples by using the intrinsic rest-frame EW drawn from Equation \ref{eqn:exp_ew} and determine the \nii/\ha~contribution as being $f_{\textrm{\nii}} = \frac{\textrm{\nii}}{\textrm{\nii} + \textrm{\ha}}$. In the case where we are modeling the intrinsic \ha~EW, we apply a $1 + f_{\textrm{\nii}}$ factor to determine the intrinsic \ha+\nii~EW. Note that narrowband surveys do not have the resolving resolution to deblend the two emission lines such that line fluxes and EWs from narrowband photometry are the combination of \ha~and \nii. Therefore, we need to incorporate an \nii/\ha~prescription to simulate narrowband and broadband photometry in our forward modeling approach.

\subsection{Mass-to-Continuum Luminosity}
\label{sec:ML_ratio}

One of the observables in narrowband surveys is the rest-frame $R$-band continuum flux density centered at 6563\AA, which is sensitive to both young and old stellar populations and traces the bulk of stellar mass in a galaxy (e.g, \citealt{Bruzual2003}). Therefore, the rest-frame $R$-band continuum observed in \ha~narrowband surveys can be used as a proxy for stellar mass.

In \citet{Khostovan2021}, we formed an empirical, single power law calibration to convert stellar mass to rest-frame $R$-band luminosity, $L_R$. This was done using COSMOS2015 stellar mass measurements \citep{Laigle2016} along with $L_R$ measured from CTIO/NB964 and DECam $z$ photometry for all sources in LAGER with a stellar mass coverage between $10^7$ and $10^{11}$ \msol~(see Figure 2 of \citealt{Khostovan2021}).

We follow a similar approach using the COSMOS2015 stellar mass catalog, cross match it with our sample catalogs, and measure the $R$-band luminosity -- stellar mass correlation. Since we are interested in the continuum luminosity and given the well-constrained photometric redshifts from the COSMOS2015 catalog, we augment the selection to include sources that have photometric redshifts within the selection range of the sample (e.g., HiZELS/NB921 selected sources with $0.35 < z_{phot} < 0.45$) but without the rest-frame \ha+\nii~EW$ > 25$\AA~(11\AA~for DAWN) and $\Sigma > 3$ criteria. This way our calibration is based on galaxies within a specific redshift range where the continuum flux density is consistent with rest-frame $R$-band centered at 6563\AA~based on the photo-$z$ regardless if emission lines are above the selection threshold or not.

\begin{figure}
	\centering
	\includegraphics[width=\columnwidth]{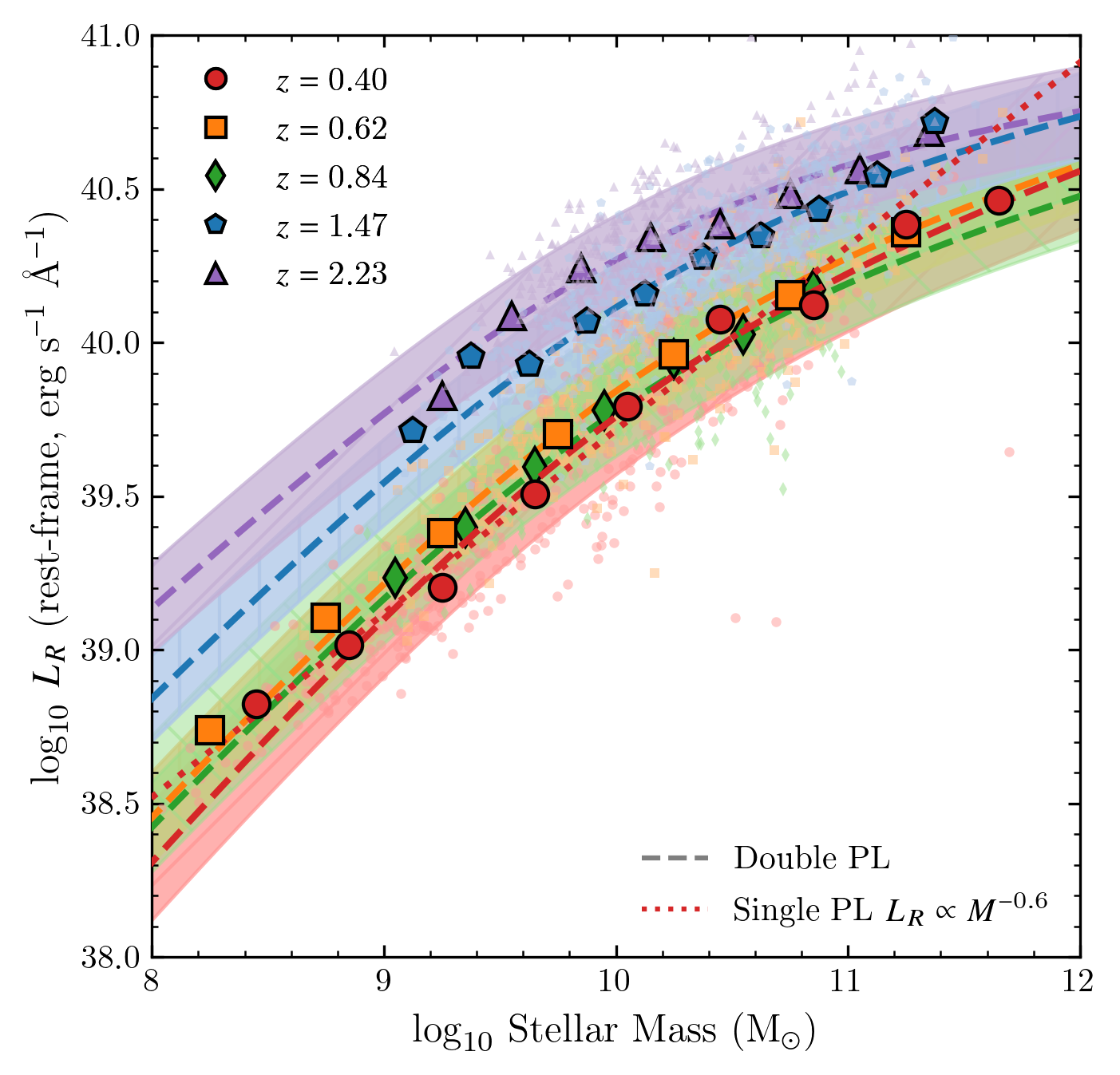}
	\caption{Mass-to-$L_R$ for all our samples with the median $L_R$ per stellar mass bin also shown as symbols. Our double power law calibration ({\it dashed} line) incorporates a pivot at $10^{10}$ \msol~where a transition in slope occurs. In comparison, we show a single power law ({\it red dashed} line) which is consistent with the low-mass end of our double power law calibration but fails to capture the high-mass end. The shaded regions is the $1\sigma$ scatter which arises from various factors, such as variations in star-formation histories and IMF. We use this calibration and the scatter to determine $L_R$  based on the assigned stellar mass for each mock \ha~emitter.}
	\label{fig:ML_ratios}
\end{figure}

Figure \ref{fig:ML_ratios} shows the $L_R$ -- stellar mass correlation for all our samples. The HiZELS/NB921 observations cover the widest range in stellar mass and can be characterized by a single power law up to $10^{10}$ \msol~ with a slope of $-0.6$ consistent with our $z = 0.47$ measurement \citep{Khostovan2021}. However, we find a shallower trend at $>10^{10}$ \msol~up to the highest stellar masses in the sample. We take this feature into account by modeling the $L_R$ -- stellar mass correlation in the form of a double power law defined as:
\begin{eqnarray}
L_R = 2 L_{R,10} \frac{y^{\gamma}}{1.+y^{-0.6}}
\label{eqn:ml_ratio}
\end{eqnarray}
where $y = M/(10^{10} \textrm{\msol}$), $\gamma$ is the high-mass end slope, and $L_{R,10}$ is the rest-frame continuum luminosity at $10^{10}$ \msol. The low-mass end is fixed to a slope of $-0.6$ constrained by the HiZELS/NB921 and LAGER/NB964 \citep{Khostovan2021} samples. This is due to the lack of coverage in the $z > 1$ samples. 

\begin{table}
	\centering
		\caption{Best-fit parameters of our mass-to-$L_R$ conversion for all our redshift slices. We assume a double power law model as defined in Equation \ref{eqn:ml_ratio} where $L_{R,10}$ is the rest-frame $R$-band continuum luminosity at $10^{10}$ \msol~with a high-mass slope $\gamma$ and a fixed low-mass slope of $-0.6$ as constrained by $z = 0.4 - 0.5$ \ha~emitters. The 1$\sigma$ scatter in $L_R$ at fixed stellar mass is also shown and used when generating mock \ha~emitters to account for variations in star-formation histories and IMFs.}
	\label{table:ml_ratio}
	{\renewcommand{\arraystretch}{1.3}
	\begin{tabular*}{\columnwidth}{@{\extracolsep{\fill}}lcccc}
		\hline
		Sample & $z$ & $\log_{10} L_{R,10}$ & $\gamma$ & $\sigma$ \\
			   &     & (erg s$^{-1}$ \AA$^{-1}$) & & (dex) \\
		\hline
	HiZELS NB921 & $0.40$ & $39.76\pm0.01$ & $0.262\pm0.013$ & $0.19$\\
	DAWN NB1066 & $0.62$ & $39.84\pm0.01$ & $0.231\pm0.013$ & $0.15$\\
	HiZELS NBJ & $0.84$ & $39.93\pm0.01$ & $0.213\pm0.021$ & $0.15$\\
	HiZELS NBH & $1.47$ & $40.11\pm0.01$ & $0.174\pm0.011$ & $0.14$\\
	HiZELS NBK & $2.23$ & $40.27\pm0.01$ & $0.104\pm0.014$ & $0.14$\\
		\hline
	\end{tabular*}
	}
\end{table}

We use Equation \ref{eqn:ml_ratio} along with the best-fit parameters shown in Table \ref{table:ml_ratio} to convert stellar mass to rest-frame $L_R$ when generating mock \ha~emitters in our forward modeling approach. We randomly perturb the assigned $L_R$ per each generated \ha~emitter by drawing from a normal distribution centered at 0 and with $1\sigma$ set to the scatter in $L_R$ at fixed stellar mass. The scatter takes into account variations in star-formation histories and IMFs per galaxy.

\subsection{Filter Profile Bias and Modeled Photometry}
\label{sec:filter_effect}

Measurements of line flux and EW in narrowband surveys assume the filter is a top-hat with 100 percent transmission and width corresponding to the narrowband FWHM. Given that narrowband filters are not perfect top-hats, this introduces a bias in the final line flux and EW measurement where intrinsically bright sources detected towards the filter wings will appear faint with low EWs. This also introduces a probed volume bias where faint galaxies will only be detected towards the filter center where transmission is at its highest resulting in low comoving volumes probed compared to intrinsically bright galaxies that can be observed even towards the filter wings (e.g., \citealt{Sobral2013,Khostovan2015,Matthee2017}).

To take this effect into account, we randomly assign a redshift to each mock emitter that covers the full wavelength coverage of the narrowband filter. We then convolve the emission line flux (measured from assigned EW$_0$ and stellar mass with mass-to-$L_R$ conversion) and continuum flux density (measured from assigned stellar mass and Equation \ref{eqn:ml_ratio}) with the narrowband and broadband transmission curves and measure the observed flux densities and magnitudes. The latter will be used in \S\ref{sec:sel_comp} when applying selection criteria. We then follow the top-hat assumption and measure the mock line fluxes, EW, and continuum flux density by assuming:
\begin{eqnarray}
	\centering
	f_\textrm{NB} = & f_c + F_L/\Delta \textrm{NB} \nonumber \\
	f_\textrm{BB} = & f_c + F_L/\Delta \textrm{BB}
\end{eqnarray}
where $f_\textrm{NB}$ and $f_\textrm{NB}$ are the flux densities observed in the narrowband and broadband filters with their corresponding FWHM defined as $\Delta \textrm{NB}$ and $\Delta \textrm{BB}$, respectively. Both flux densities are a combination of the observed continuum flux density, $f_c$, and line flux $F_L$. We use this $F_L$ and rest-frame EW ($F_L/(f_c (1+z))$) affected by the top-hat assumption to compare with observations and constrain EW distributions.

\subsection{Constraining EW distributions}
\label{sec:sel_comp}

The final step to form a mock \ha~sample is to apply the selection criteria corresponding to the observed sample that we want to compare to and constrain the EW distributions. This is done by using the mock photometry generated using the mock line flux, continuum flux density, and EW convolved with the narrowband and broadband transmission curves as described in \S\ref{sec:filter_effect}. The selection criteria for each of our observed samples is described in \S\ref{sec:samples} and is primarily a combination of an EW cut, line flux cut, and color significance cut. Since the narrowband surveys in this work also have inhomogeneity in depth and survey area as shown in Figure \ref{fig:HiZELS_NBH}, we apply the selection criteria on a pointing-to-pointing basis.

We use the mock \ha~samples to constrain the intrinsic EW distributions by subdividing the sample in bins of stellar mass and comparing the underlying EW distributions of both the observed and mock samples (with selection effects applied to mimic the observations). Since the mock samples are dependent on the intrinsic \wo~that was assigned to the sample, we know then based on our assumptions what the intrinsic EW distribution should be to reproduce the EW distributions in observations when selection biases are included. 

We start by normalizing both distributions (observed sample and mock) to unity, where the binning scheme is based on the EW distribution from the observed sample. We then follow a $\chi^2$ minimization approach:
\begin{eqnarray}
\centering
\chi^2 = \sum_{i=0}^{i=N} \bigg (\frac{n_{o,i} - n_{m,i}(\textrm{\wo})}{\sigma_{o,i}}\bigg)^2
\end{eqnarray}
where $n_{o,i}$ and $n_{m,i}$ are the normalized number of emitters in the observed and mock samples for the $i^\textrm{th}$ bin, respectively, and $\sigma_{o,i}$ is the Poisson error for the observed sample in the $i^\textrm{th}$ bin. We report the assigned intrinsic \wo~that minimizes $\chi^2$ as the best-fit that describes the intrinsic \ha~distribution that best matches the EW distribution in observations when selection biases are applied.

\section{Results}
\label{sec:results}

\subsection{H$\alpha$ Equivalent Width and Continuum/Stellar Mass}
\label{sec:EW_stellar_mass}
\begin{figure*}
	\centering
	\includegraphics[width=\textwidth]{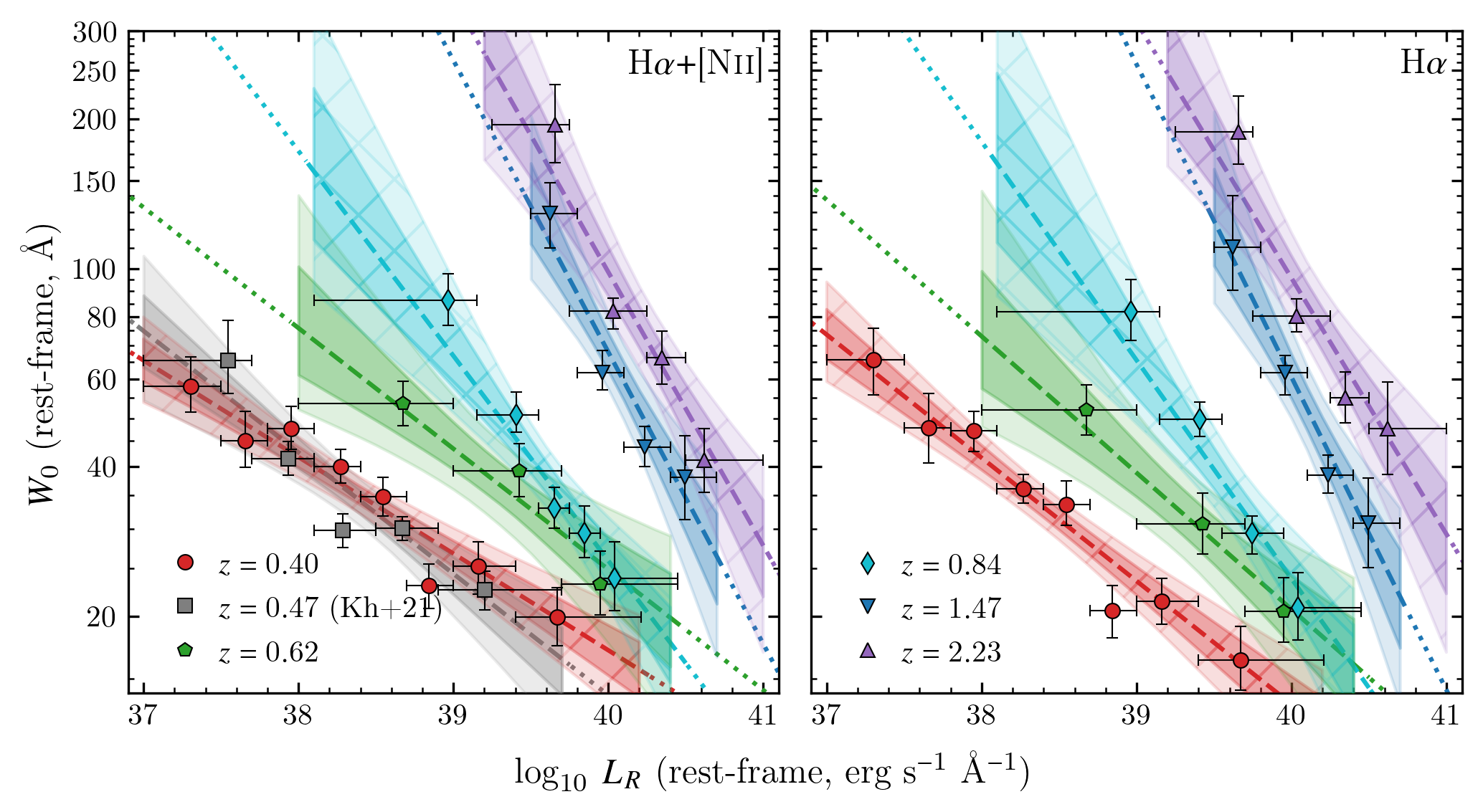}
	\caption{The intrinsic correlation between Equivalent Width and rest-frame $R$-band Luminosity with the {\it left} panel showing the case for \ha$+$\nii~and the {\it right} panel for \ha. In both cases, we find a strong anti-correlation between \wo~and $L_R$ such that faint continuum sources have higher \wo(\ha$+$\nii) and \wo(\ha). We also find \wo~increases by an order-of-magnitude with redshift at fixed $L_R$ signifying a strong redshift evolution. Included in the {\it left} panel are the \wo(\ha$+$\nii) measurements from our previous work using the LAGER \ha~sample \citep{Khostovan2021} where we find strong agreement with our $z = 0.4$ \wo~measurements.}
	\label{fig:ha_n2_w0_cont}
\end{figure*}

\begin{figure*}
	\centering
	\includegraphics[width=\textwidth]{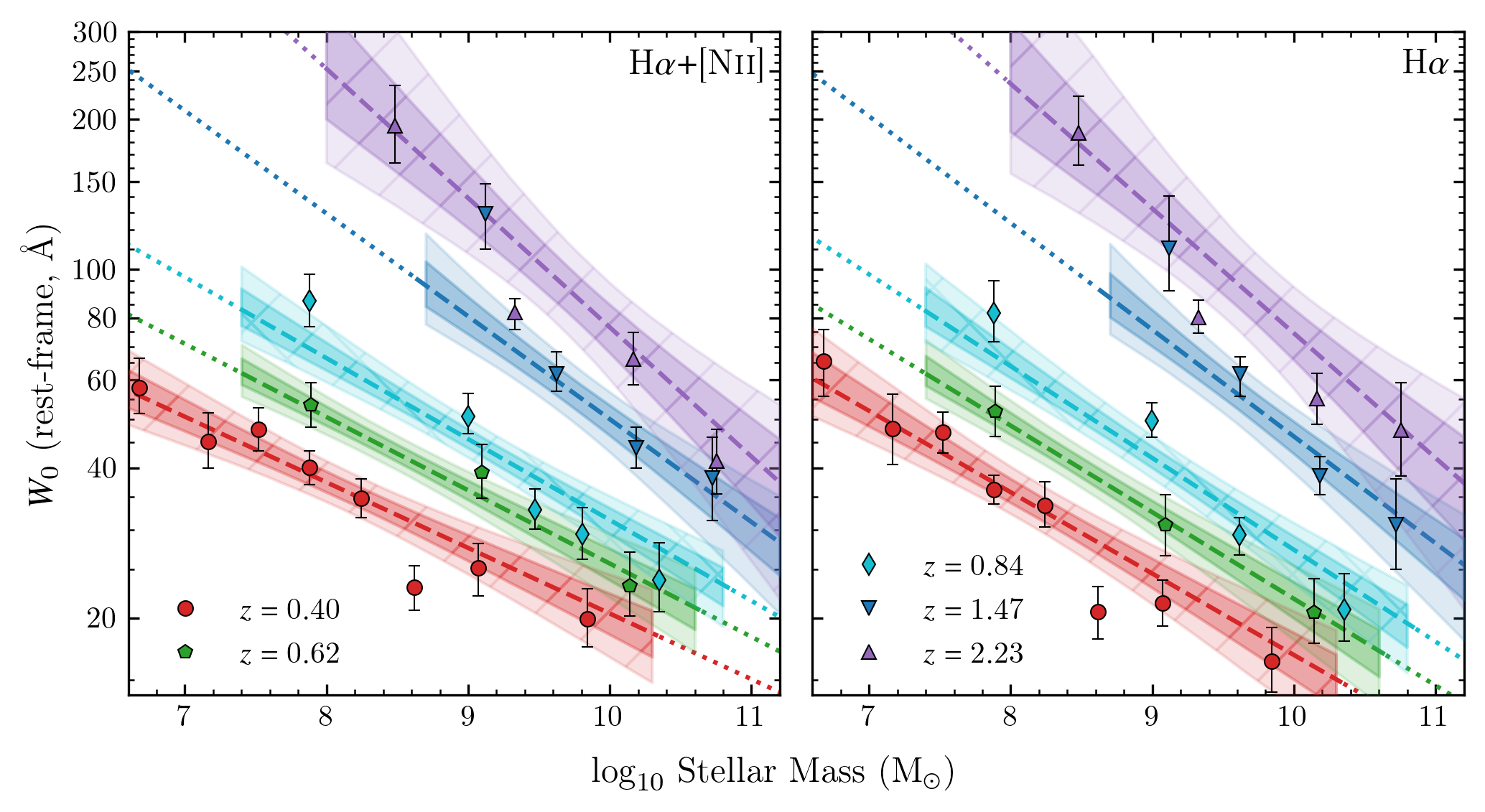}
	\caption{The intrinsic correlation between Equivalent Width and Stellar Mass where the panel descriptions are the same as Figure \ref{fig:ha_n2_w0_cont}. Our best-fit \wom~model, as described in Equation \ref{eqn:EW_model}, is shown in each panel as a {\it dashed} line with the $1\sigma$ and $2\sigma$ confidence regions shown as {\it dark} and {\it light} colored regions, respectively. The {\it dotted} lines are an extrapolation of our \wom~model for where we have no observational constraints. We find a strong redshift evolution where low-mass galaxies have a steeper increase in \wo~with increasing redshift compared to high-mass galaxies.}
	\label{fig:ha_n2_w0_mass}
\end{figure*}

\begin{table*}
	\centering
	\caption{The \ha~and \ha$+$\nii~Equivalent Width anti-correlation with rest-frame continuum $L_R$ luminosity (centered at 6563\AA) and stellar mass. \wo$_{40}(z)$ and \wo$_{10}(z)$ correspond to the typical rest-frame \ha~EW at $L_R = 10^{40}$ erg s$^{-1}$ \AA$^{-1}$ (rest-frame) and $10^{10}$ \msol, respectively. $\beta(z)$ and $\gamma(z)$ are the corresponding slopes of the \wo~correlation with $L_R$ and stellar mass, respectively, and are defined based on Equations \ref{eqn:powerlaw_cont} and \ref{eqn:powerlaw_mass}.}	\label{table:EW_correlation}
	{\renewcommand{\arraystretch}{1.3}
	\begin{tabular*}{\textwidth}{@{\extracolsep{\fill}} lccccc}
		\hline
			& & \multicolumn{2}{c}{$L_R$ Correlation} & \multicolumn{2}{c}{Stellar Mass Correlation} \\
		Line & $z$ & $W_{0,40}(z)$ & $\beta(z)$ & $W_{0,10}(z)$ & $\gamma(z)$ \\
			&  & (\AA) & & (\AA) & \\
		\hline
		H$\alpha$+[N{\sc ii}] & $0.40$ & $17.09_{-1.88}^{+2.10}$ & $-0.19_{-0.03}^{+0.03}$ & $18.77_{-2.41}^{+2.57}$ & $-0.14_{-0.03}^{+0.03}$ \\
		& $0.62$ & $24.85_{-3.69}^{+4.09}$ & $-0.24_{-0.09}^{+0.07}$ & $27.03_{-4.56}^{+4.51}$ & $-0.15_{-0.06}^{+0.04}$ \\
		& $0.84$ & $25.87_{-2.56}^{+2.74}$ & $-0.41_{-0.10}^{+0.09}$ & $28.66_{-3.47}^{+3.85}$ & $-0.22_{-0.06}^{+0.04}$ \\
		& $1.47$ & $67.93_{-5.61}^{+6.38}$ & $-0.59_{-0.14}^{+0.12}$ & $57.40_{-7.28}^{+7.57}$ & $-0.29_{-0.11}^{+0.08}$ \\
		& $2.23$ & $98.36_{-10.20}^{+12.22}$ & $-0.55_{-0.12}^{+0.10}$ & $68.57_{-10.72}^{+9.40}$ & $-0.25_{-0.08}^{+0.06}$ \\
		\hline
		H$\alpha$ & $0.40$ & $13.32_{-1.63}^{+1.83}$ & $-0.25_{-0.03}^{+0.03}$ & $15.01_{-2.34}^{+2.42}$ & $-0.18_{-0.03}^{+0.03}$ \\
		& $0.62$ & $20.66_{-3.19}^{+3.43}$ & $-0.27_{-0.09}^{+0.07}$ & $22.75_{-4.44}^{+4.07}$ & $-0.17_{-0.07}^{+0.05}$ \\
		& $0.84$ & $23.81_{-2.89}^{+2.95}$ & $-0.44_{-0.11}^{+0.10}$ & $26.44_{-4.24}^{+3.80}$ & $-0.23_{-0.07}^{+0.05}$ \\
		& $1.47$ & $60.48_{-5.75}^{+6.75}$ & $-0.63_{-0.16}^{+0.14}$ & $50.29_{-6.94}^{+8.05}$ & $-0.31_{-0.12}^{+0.08}$ \\
		& $2.23$ & $95.67_{-8.92}^{+10.45}$ & $-0.51_{-0.12}^{+0.11}$ & $67.67_{-9.98}^{+10.45}$ & $-0.24_{-0.08}^{+0.06}$ \\
		\hline
	\end{tabular*}
	}
\end{table*}

Past studies of star-forming galaxies have reported a \ha~EW -- stellar mass anti-correlation with the typical EW increasing as $(1+z)^{1.8}$ up to $z \sim 2$ at fixed stellar mass (e.g., \citealt{Fumagalli2012,Sobral2014,Faisst2016}). However, we showed in \citet{Khostovan2021} that selection limits can severely bias the underlying correlation by increasing the EW -- stellar mass slope by a factor of $\sim 1.6$ based on the $z \sim 0.5$ \ha~LAGER. In this section, we extend that work to investigate the \ha~EW -- continuum $R$-band luminosity/stellar mass anti-correlation where we follow the methodology outlined in \S\ref{sec:modeling} to measure the `intrinsic' anti-correlation (selection/sample bias corrected).

Figure \ref{fig:ha_n2_w0_cont} shows EW$_0$(\ha$+$\nii) and EW$_0$(\ha) in terms of rest-frame $L_R$ continuum luminosity centered at $6563$\AA. We find a clear EW -- $L_R$ anti-correlation at all observed redshifts with the slope becoming steeper with increasing redshift. At fixed $L_R$, the typical EW increases with redshift highlight a strong redshift evolution. For each sample, we fit a simple power law of the form:
\begin{eqnarray}
	\textrm{\wo}(L_R) = W_{0,40} \Bigg(\frac{L_R}{10^{40} \textrm{erg s}^{-1} \textrm{cm}^{-2} \textrm{\AA}^{-1}}\Bigg)^\beta [\textrm{\AA}]
	\label{eqn:powerlaw_cont}
\end{eqnarray}
where \wo~is the typical rest-frame EW at a $L_R$, $\beta$ is the correlation slope, $W_{0,40}$ is the typical rest-frame EW at $L_R = 10^{40}$ erg s$^{-1}$ cm$^{-2}$ \AA$^{-1}$. The best-fits are overlaid in Figure \ref{fig:ha_n2_w0_cont} with the $1\sigma$ and $2\sigma$ confidence regions highlighted as {\it dark} and {\it light} colored regions. The best-fit parameters are also shown in Table \ref{table:EW_correlation}.

The HiZELS samples show an EW(\ha+\nii) -- $L_R$ anti-correlation with statistical significance $>4\sigma$ from a null hypothesis (no correlation), while the DAWN $z = 0.62$ sample shows $\sim 3\sigma$ significance which is mostly due to the smaller sample size in comparison to the HiZELS samples. The slope is found to become steeper with increasing redshift from $\beta(z=0.4) = -0.19$ to $\beta(z = 2.23) = -0.55$ corresponding to a factor of $\sim 3$ increase. The \ha~EW -- $L_R$ anti-correlation also shows a strong statistical significance of $\sim 4 - 8\sigma$ from a null hypothesis with a steeper slope from $\beta(z = 0.4) = -0.25$ to $\beta(z = 2.23) = -0.51$ (a factor of $\sim 2$ increase). The shallower increase in the slope compared to the EW(\ha+\nii) -- $L_R$ anti-correlation is attributed to the contribution of \nii~at low EWs where \nii/\ha~is higher.

The strong increase in the slope for both EW(\ha+\nii) and EW(\ha) anti-correlations with $L_R$ is somewhat driven by the changing $M/L_R$ with redshift as shown in \S\ref{sec:ML_ratio} and Figure \ref{fig:ML_ratios}. We take this into account by repeating the same measurement and analysis by converting $L_R$ to stellar mass using our $M/L_R$ calibration from Equation \ref{eqn:ml_ratio}. Figure \ref{fig:ha_n2_w0_mass} shows the anti-correlation with stellar mass where we fit simple power laws of the form:
\begin{eqnarray}
		\textrm{\wo}(M) = W_{0,10} \Bigg(\frac{M}{10^{10} \textrm{M}_\odot} \Bigg)^\gamma [\textrm{\AA}]
		\label{eqn:powerlaw_mass}
\end{eqnarray}
where $W_{0,10}$ is the typical EW for an emitter at 10$^{10}$ \msol~and $\gamma$ is the anti-correlation slope. The best-fit power laws are shown in Table \ref{table:EW_correlation} where we find that the slopes become shallower in comparison to the EW--$L_R$ for both EW(\ha+\nii) and EW(\ha). We find that the anti-correlation between EW and stellar mass for both emission lines has $>3.5\sigma$ significance. The slope becomes steeper with increasing redshift although shallower in comparison to the EW--$L_R$ anti-correlation. This implies that the previous steepness is primarily driven by changes in $M/L_R$ over cosmic time; however, a steeper slope in the EW--stellar mass anti-correlation is still present. This suggests that the EW distributions of low-mass \ha~emitter populations become wider and consist of even higher EW systems with increasing redshifts. It also implies that the contrast in EW distributions between low-mass and high-mass \ha~systems becomes even stronger with increasing redshift as low-mass systems have wider EW distributions than high-mass systems by cosmic noon.

\subsubsection{Redshift Evolution of Typical \ha~Equivalent Widths}

The typical \ha~EW for 10$^{10}$ \msol~\ha~emitters is also seen to increase with redshift rising from $\sim 15$\AA~at $z \sim 0.4$ to $\sim70$\AA~by $z \sim 2$ suggesting a redshift evolution where \ha~emitters at high-$z$ have significantly higher EWs compared to their low-$z$ counterparts. Figure \ref{fig:ha_n2_w0_mass} shows that at any fixed stellar mass the typical EW increases with increasing redshift but also shows that the redshift evolution is mass dependent where low-mass \ha~emitters have a stronger increase in \wo~with increasing redshift.

We model this mass-dependent redshift evolution using a power law of the form:
\begin{eqnarray}
	W_0(M,z) = W_{0,10}(1+z)^{\alpha_1} \Bigg(\frac{M}{10^{10} M_\odot}\Bigg)^{\gamma_{0} + \alpha_2(1+z)} [\textrm{\AA}]
	\label{eqn:EW_model}
\end{eqnarray}
which is a combination of two main properties: 1) the evolution of \wo~highlighted by $W_{0,10}$ (typical EW of a $z = 0$ \ha~emitter with stellar mass of $10^{10}$ \msol) and a redshift evolution defined by the slope, $\alpha_1$; and, 2) the evolution in the anti-correlation slope described by $\gamma_0$ (power law slope at $z = 0$) that evolves with redshift with a slope, $\alpha_2$. We simultaneously fit the EW--stellar mass anti-correlation and redshift evolution using all our samples and \wo($M$) measurements. 

Figure \ref{fig:ha_n2_w0_mass} shows the best-fits of our redshift and stellar mass dependent EW model with the parameters highlighted in Table \ref{table:best_model}. We find that the typical \ha~EW at $10^{10}$ \msol~evolves as $\sim (1+z)^{1.8}$, similar to previous measurements up to $z \sim 2$ (e.g. \citealt{Fumagalli2012,Sobral2014,Faisst2016}). At $z = 0$, the \ha+\nii~EW -- stellar mass anti-correlation is nearly flat with $\gamma_0 = - 0.03^{+0.07}_{-0.06}$ while the \ha~EW anti-correlation has a slope of $-0.10^{+0.07}_{-0.07}$, which is probably due to an increase in \nii~contribution at the high-mass end causing a shallower slope. However, both the EW(\ha+\nii) and EW(\ha) show a steepening slope where $\gamma(z \sim 2)$ is $\sim -0.25$ and $\sim -0.35$ for \ha+\nii~and \ha, respectively. This implies that \ha~emitters not only have increasingly higher typical EWs with increasing redshift, but also their low-mass populations show increasingly wider EW distributions in comparison to their high-mass counterparts. In the scope of star-formation activity, our results suggest that populations of low-mass \ha~emitters have a wide, diverse range of star-forming galaxies from significantly high EWs corresponding to high sSFR (active, bursty star-forming systems) to low EW, steady star-formation histories.

\begin{table}
	\centering
	\caption{The best-fit properties of our redshift and stellar-mass dependent equivalent width model as defined in Equation \ref{eqn:EW_model} for both \ha$+$\nii~and \ha~emission. $W_{0,10}$ is the median EW at $z = 0$ for $10^{10}$ \msol~star-forming galaxies with a $(1+z)^{\alpha_1}$ evolution. $\gamma_0$ is the \wo--stellar mass anti-correlation slope at $z = 0$ and evolves as $\alpha_2(1+z)$.}
	\label{table:best_model}
	{\renewcommand{\arraystretch}{1.3}
	\begin{tabular*}{\columnwidth}{@{\extracolsep{\fill}} lcccc}
		\hline
		Line & $W_{0,10}$ & $\alpha_1$ & $\gamma_0$ & $\alpha_2$\\
		& (\AA) & & & \\
		\hline
		H$\alpha$+[N{\sc ii}] & $11.98_{-1.75}^{+2.04}$ & $1.58_{-0.21}^{+0.20}$ & $-0.03_{-0.06}^{+0.07}$ & $-0.07_{-0.04}^{+0.03}$\\
		H$\alpha$ & $9.26_{-1.53}^{+1.71}$ & $1.78_{-0.23}^{+0.22}$ & $-0.10_{-0.07}^{+0.07}$ & $-0.05_{-0.04}^{+0.03}$\\
		\hline
	\end{tabular*}
	}
\end{table}

\subsubsection{A Biased View of the \ha~EW Evolution}

\begin{figure}
	\centering
	\includegraphics[width=\columnwidth]{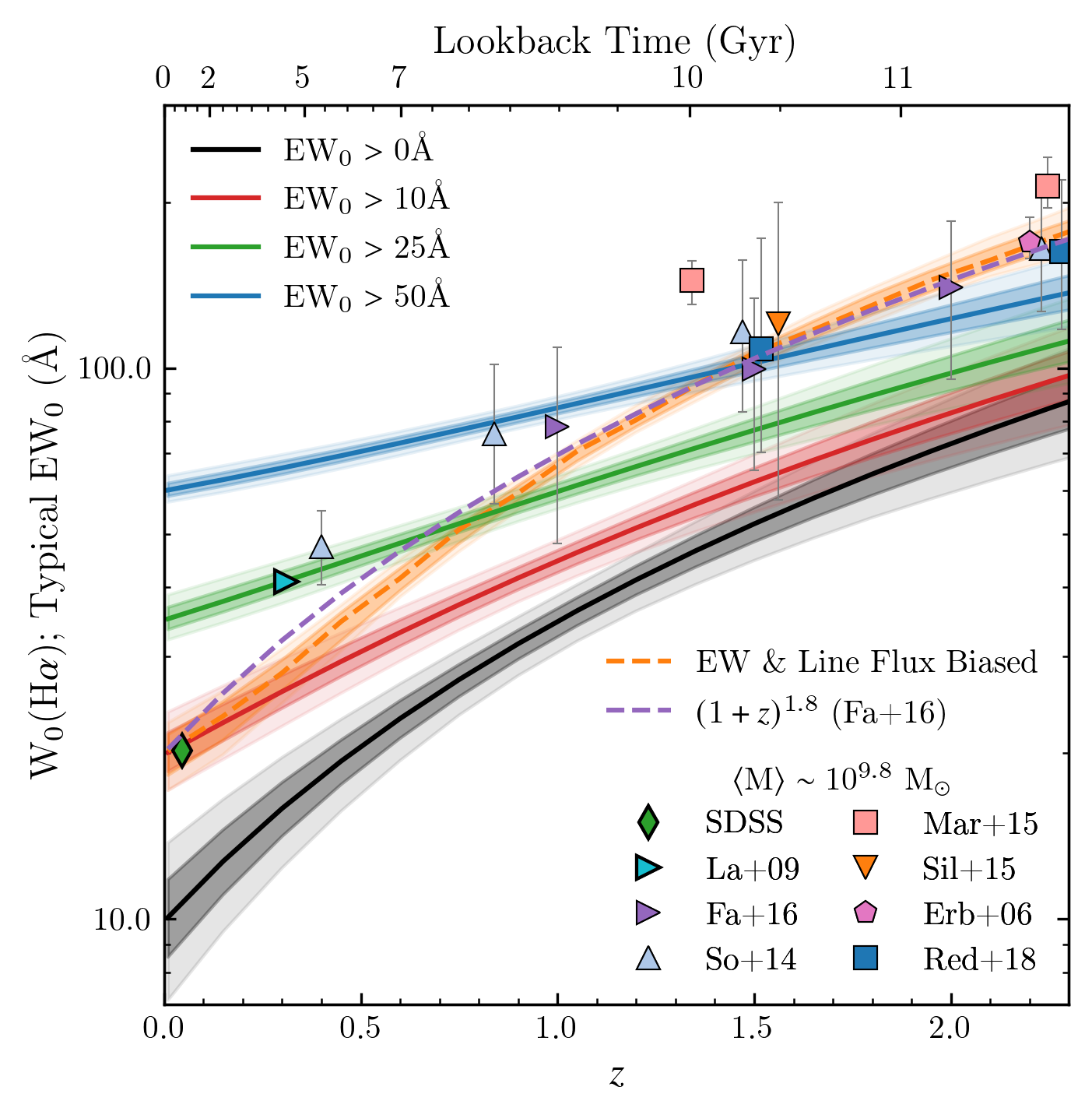}
	\caption{Redshift Evolution of Typical \ha~rest-frame EW with stellar masses between $10^{9.5}$ \msol~and $10^{10}$ \msol. We find that the intrinsic \ha~EW increases in this stellar mass range by an order of magnitude from the local Universe to $z \sim 2$ encompassing 11 Gyr of cosmic history. Included are measurements of \ha~EWs at various cosmic epochs \citep{Erb2006,Lamareille2009,Fumagalli2012,Sobral2014,Silverman2015,Faisst2016,Marmol2016,Reddy2018} which are found to be well above our measured intrinsic EW evolution. However, applying an EW$_0 > 10$\AA~and \ha~line flux cut $> 10^{-16.5}$ erg s$^{-1}$ cm$^{-2}$ can explain the discrepancy between our intrinsic measurement and reported observations. This signifies how selection is an important factor in constraining the redshift evolution of \ha~EWs and suggests past measurements are biased by a normalizing factor due to selection.}
	\label{fig:EW_zevol}
\end{figure}

In our previous work \citep{Khostovan2021}, we showed how selection biases can result in a steeper EW -- stellar mass correlation for the $z = 0.47$ \ha~sample drawn from the narrowband LAGER survey. Figure \ref{fig:EW_zevol} shows the redshift evolution of typical \ha~rest-frame EW at a fixed stellar mass range of $10^{9.5 - 10}$ \msol~for different selection limit cases. We find that typical \ha~EW corrected for selection bias (\textit{black solid line}) increases by nearly an order of magnitude from $z \sim 0$ to $z \sim 2$ and is consistent with a $(1+z)^{1.8}$ evolution found in previous studies (e.g., \citealt{Faisst2016}). However, we find observations to have systematically higher measurements of \ha~EWs up to $z \sim 2$. 

We test how selection limits alone can resolve the discrepancy between past observations and our intrinsic measurement of the \ha~EW redshift evolution. In a typical narrowband survey, two major selection limits are applied to identify potential emission line galaxy candidates. The first is an EW cut to remove bright narrowband sources that have strong continuum features mimicking emission lines in narrowband photometry. The second is a line flux cut (e.g., narrowband magnitude limit) to remove sources that are close to a certain depth in the narrowband images.

Figure \ref{fig:EW_zevol} shows the typical evolution for varying rest-frame EW cuts and we find an increasing typical EW measurement with increasing selection limit at fixed redshift. The effect is more pronounced at lower redshift and this is caused by the narrower EW distributions at such regimes (e.g., low \wo). Applying a line flux limit of $10^{-16.5}$ erg s$^{-1}$ cm$^{-2}$ in conjunction with an EW$_0 > 10$\AA~limit strongly matches with the $(1+z)^{1.8}$ evolution of \citet{Faisst2016} and observational measurements. This does not imply each survey has these specific selection limits as other combinations of line flux and EW$_0$ limits can match with the observations. The main point is that it clearly shows not correcting for selection limits will result in an inflated typical \ha~EW measurement for any stellar mass range. This highlights the importance of measuring the intrinsic \ha~EW distributions, its anti-correlation with stellar mass, and its redshift evolution. 

\begin{figure*}
	\centering
	\includegraphics[width=\textwidth]{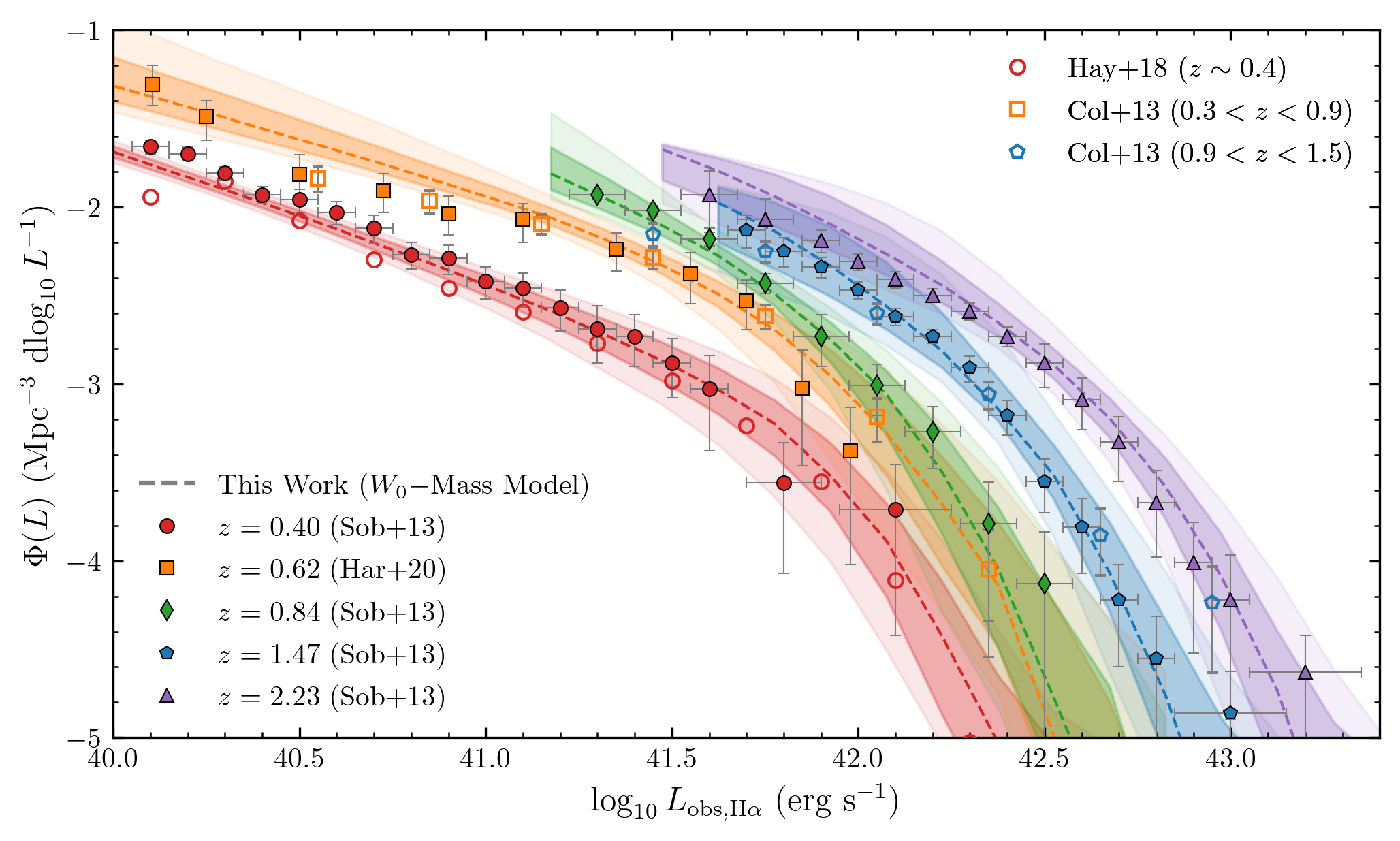}
	\caption{The observed (not corrected for dust) \ha~luminosity function based on our best-fit \wom~model. Included are the observed \ha~LFs from \citet{Sobral2013} and \citet{Harish2020} as filled symbols where we find a strong agreement for all redshift slices. We also include luminosity functions from HyperSuprimeCam at $z = 0.4$ \citet{Hayashi2018} and the {\it HST}/WISP grism survey of \citet{Colbert2013} where we also find strong agreement with our \ha~LFs. This suggests that our `intrinsic' \wom~model can produce \ha~populations consistent with observations and describes both the \ha~EW distributions and LFs coupled with a stellar mass function.}
	\label{fig:predicted_LFs}
\end{figure*}

\subsection{Equivalent Width and \ha~luminosity functions}
\label{sec:HA_LFs}
Three major statistical properties of an emission line galaxy population are the stellar mass function (integrated past star-formation history), equivalent width distribution (the ratio between instantaneous and past star formation activity), and the luminosity function (instantaneous star formation activity). Our forward modeling approach described in \S\ref{sec:approach} assumes a stellar mass function and $M/L_R$ to populate our samples with \ha~emitters with varying luminosities and EWs. We constrain the `intrinsic' EW distribution and show that it depends strongly on both redshift and stellar mass (see Figures \ref{fig:ha_n2_w0_cont} and \ref{fig:ha_n2_w0_mass}). In this section, we test if our \wo$(M,z)$ model can also reproduce \ha~luminosity functions which would provide strong evidence supporting our model and approach.

We follow a similar approach as discussed in \S\ref{sec:approach} with the difference being that we assign EW to our samples based on the best-fit \wo$(M,z)$ in Equation \ref{eqn:EW_model} and Table \ref{table:best_model}. \ha~luminosity is measured using the rest-frame EW and rest-frame $R$-band luminosity. We do not assume any prescription for dust in this case such that \ha~luminosities reported here are based on observered (not corrected for dust) luminosities. We also apply a constant EW cut consistent with HiZELS and DAWN when making our predicted \ha~LFs to compare with observations. This is because the observed LFs are drawn from EW-limited samples.

Figure \ref{fig:predicted_LFs} shows our predicted \ha~LFs based on assuming the \citet{Sobral2014} SMF along with the intrinsic \ha~$\textrm{\wo}(M,z)$ model that was constrained using the HiZELS and DAWN \ha~EW distributions. We find a strong agreement between our predicted LFs and the observed LFs from \citet{Sobral2014} and \citet{Harish2020}. This suggests that our observationally-constrained $\textrm{\wo}(M,z)$~model can recover the \ha~LF based on simple assumptions. However, despite this strong agreement, there are some minor discrepancies. For example, the predicted $z = 0.40$ LF is slightly underestimated at $L_\textrm{\ha} < 10^{40.75}$ erg s$^{-1}$. We also find the predicted $z = 2.23$ \ha~LF somewhat overestimates in the faint-end although still within 1$\sigma$. The strong agreement with observed LFs is an indication that our `intrinsic' EW--stellar mass--$z$ model can robustly describe all three major statistical properties of emission line galaxies. In our previous work, we found similar agreement in \citet{Khostovan2021} where the `intrinsic' \wo--stellar mass anti-correlation was able to recover the \ha~LF while not correcting for selection biases resulted in an underestimation of the bright-end.  We also find that our predicted \ha~LFs are in strong agreement with the $z = 0.4$ \ha~LF of \citet{Hayashi2018}, which is based on a 16 deg$^2$ survey using Subaru/HSC. We also find a strong agreement with the \ha~LFs from the {\it HST} grism survey, WISP \citep{Colbert2013}, suggesting that our approach can also describe \ha~populations regardless of selection. 

Overall, the agreement between our predicted \ha~LFs and the observed LFs from the literature gives strong evidence that our intrinsic \wo($M$,$z$) model empirically constrained by the HiZELS and DAWN samples can robustly describe the general \ha~emission line galaxy population up to cosmic noon. It also allows us to investigate global \ha~population properties at different equivalent widths (sSFRs) which, in turn, allows us to study the variety of star-formation histories and how it contributes to the total cosmic star-formation activity at various epochs.

\subsection{\ha~Star Formation Rate Functions}
\label{sec:HA_SFRFs}
Following the same approach used in \S\ref{sec:HA_LFs}, we use our best-fit \wom~model to investigate how well it can recover the dust-corrected \ha~star formation rate functions (SFRFs). We begin by generating mock \ha~samples using our best-fit \wom~model with observed \ha~luminosity measured by using the rest-frame EW and $R$-band continuum luminosity. This is then converted to star formation rates using the \citet{Kennicutt1998} calibration
\begin{equation}
	\centering
	\textrm{SFR}(\textrm{\ha}) = 4.4\times10^{42} L_\textrm{\ha} ~\textrm{\msol} ~\textrm{yr}^{-1}
	\label{eqn:SFR_calibration}
\end{equation}
where we assume a \citet{Chabrier2003} IMF. Dust corrections are then applied to each mock \ha~emitter by using the \citet{Garn2010} \aha--stellar mass prescription, which is calibrated on SDSS DR7 but has been shown to work up to $z \sim 2$ (e.g., \citealt{Dominguez2013,Kashino2013,Sobral2016,Shapley2022}). For each mock emitter we assign \aha~using a normal distribution, $\mathcal{N} (\mu=\textrm{\aha}(M),\sigma=0.28~\textrm{mag})$, which is centered on \aha~defined by the calibration at a given stellar mass and perturbed by a random number drawn from a normal distribution with $\sigma = 0.28$ mag to incorporate the observational scatter in the dust correction \citep{Garn2010}.

\begin{figure}
	\centering
	\includegraphics[width=\columnwidth]{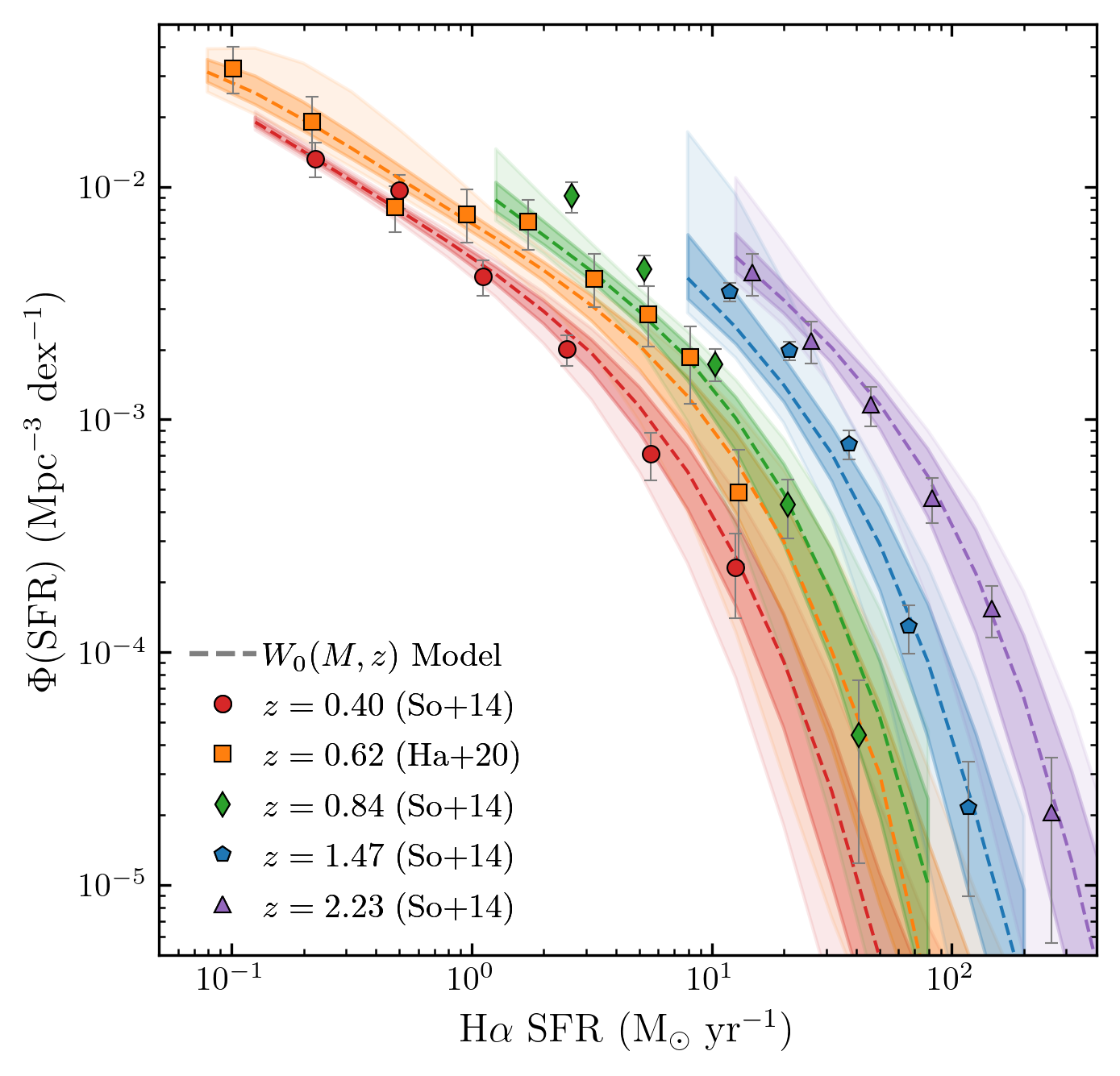}
	\caption{The \ha~star formation rate functions (SFRFs) based on our best-fit \wom~model assuming the \citet{Kennicutt1998} \ha~calibration and stellar mass--dependent dust correction of \citet{Garn2010}. We compare to the \citet{Sobral2014} SFRFs and infer the $z = 0.62$ SFRF of \citet{Harish2020} by using their \citet{Garn2010} dust corrected \ha~LFs and the \citet{Kennicutt1998} calibration. We find a strong agreement between our SFRFs and those of the literature which implies that our \wom~model can predict the SF distribution of \ha~emitters up to $z \sim 2$.}
	\label{fig:SFRFs}
\end{figure}

Figure \ref{fig:SFRFs} shows the SFRFs assuming our \wom~model compared with the \ha~star formation rate functions (SFRFs) from \citet{Sobral2014} and \citet{Harish2020}. We find a strong agreement between our predicted SFRFs and those from the literature at all redshift slices, specifically for the $z = 0.4$ and $z = 2.23$ SFRFs where our predicted SFRFs are in near perfect agreement with observations. Our $z = 1.47$ SFRF slighlty underestimates the number densities $<40$ \msol~yr$^{-1}$ although it is still within $1\sigma$ uncertainty. The main difference occurs with the $z = 0.84$ SFRF where we find our predicted SFRF underestimates the observed number densities at $<6$ \msol~yr$^{-1}$. Overall, our \wom~model shows that we can describe both the observed \ha~LF and the dust-corrected \ha~SFRF up to $z \sim 2.2$ by using information regarding the intrinsic \ha~equivalent width distributions and an assumption of the stellar mass function for star-forming galaxies.

\subsection{Cosmic Star Formation Rates: Which types of galaxies contribute the most?}
\label{sec:results_SFRD}

\begin{figure*}
	\centering
	\includegraphics[width=\textwidth]{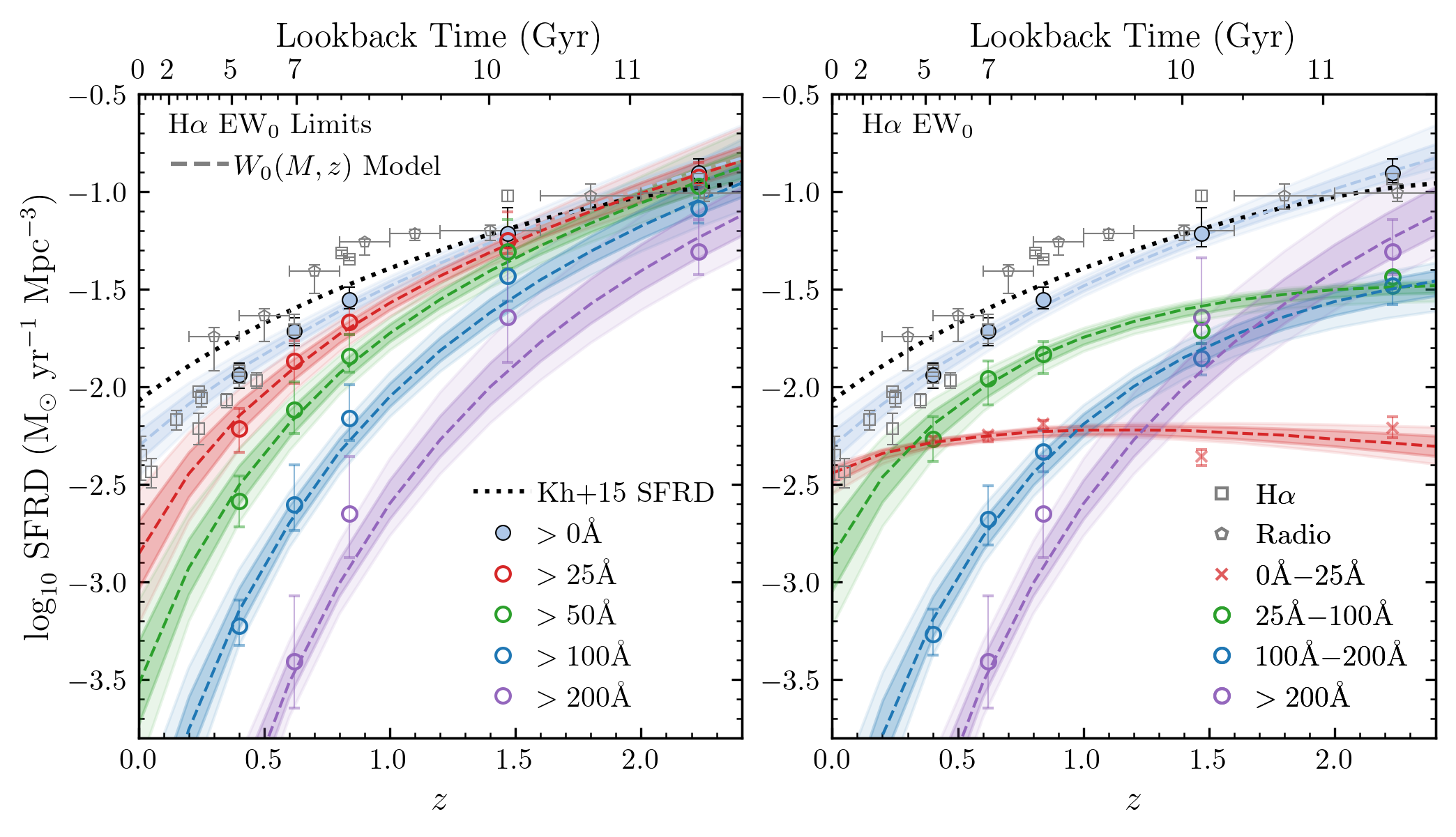}
	\caption{Cosmic star-formation rate density (SFRD) evolution for different \ewr~limits ({\it left}) and bins of \ewr ({\it right}). \ha~({\it square}; \citealt{Ly2007,Westra2010,Sobral2014,Sobral2015,Stroe2015,Harish2020,Khostovan2020,VilellaRojo2021} and radio ({\it pentagon}; \citealt{Karim2011}) SFRD measurements are also included along with the \citet{Khostovan2015} \oii~SFRD evolution. The SFRD from our \wom~model is shown as a {\it dashed line} with the full \ha~population SFRD from our model shown in {\it light blue} and is consistent with the \ha~SFRD measurements from the literature. We find that \ewr$>100$\AA~emitters contribute the most to cosmic SFRD at $z > 1.5$ and rapidly decrease in SFRD towards $z \sim 0$. Intermediate systems (\ewr~$25$ -- $100$\AA) contribute the most at $z \sim 0.4$ to $1.5$ followed by \ewr$<25$\AA~systems at $z < 0.4$. This highlights how high EW \ha~emitters are important populations in regards to overall star-formation activity in the high-$z$ Universe.}
	\label{fig:SFRD_EWs}
\end{figure*}

\begin{figure*}
	\centering
	\includegraphics[width=\textwidth]{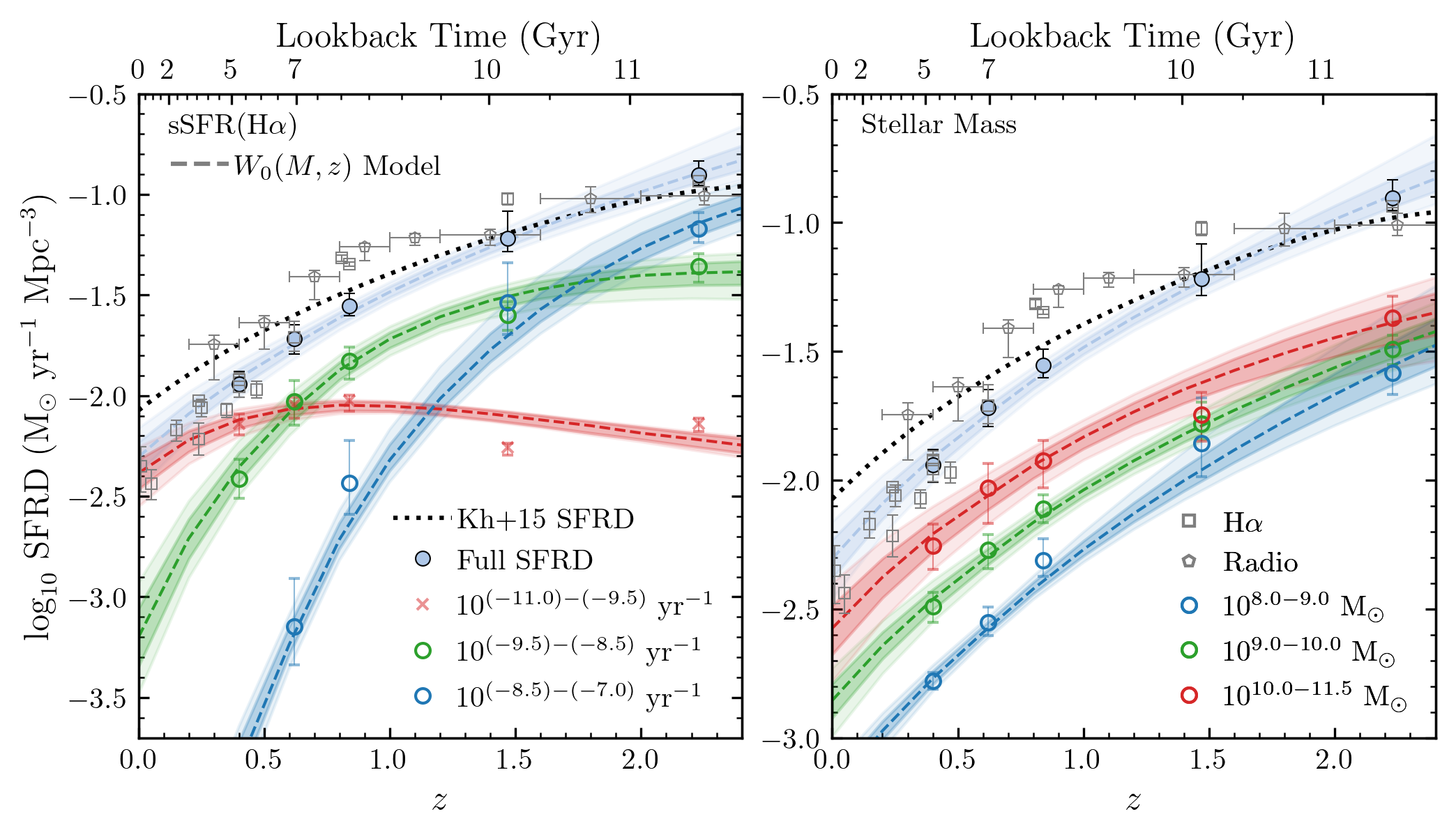}
	\caption{Literature and formatting description same as Figure \ref{fig:SFRD_EWs} except for different ranges of sSFR ({\it left panel}) and stellar mass ({\it right panel}). We find high sSFR systems ($>10^{-8.5}$ yr$^{-1}$) contribute the most to cosmic SFR at $z > 1.7$ followed by intermediate sSFRs ($10^{-9.5}$ yr$^{-1}$ to $>10^{-8.5}$ yr$^{-1}$ yr$^{-1}$) at $0.5 < z < 1.7$ and followed by $<10^{9.5}$ yr$^{-1}$ systems towards $z \sim 0$. The {\it right panel} shows massive galaxies contribute the most to overall cosmic SFR up to $z \sim 2$ but their relative contribute decreases as the gap in SFRD between low and high mass galaxies diminishes approaching $z \sim 2$. This would suggest that high sSFR and EW systems regardless of stellar mass are the main contributors to overall cosmic SFRD at $z > 1.5$.}
	\label{fig:SFRD_ssfr_mass}
\end{figure*}

Given that equivalent width is a ratio between emission line flux (instantaneous SF) and continuum flux density (average SF over longer timescales), it can be used as a proxy in understanding star-formation histories of galaxies. We have shown in \S\ref{sec:HA_LFs} and \ref{sec:HA_SFRFs} how our \wom~model can produce \ha~LFs and SFRFs consistent with observations. This allows us to use our model to investigate what types of galaxies contribute the most to overall cosmic star formation activity at different epochs in galaxy evolution by measuring the relative SFR contribution of star-forming galaxies at different cosmic times in varying bins of \ha~EW and threshold limits, stellar mass, and sSFR(\ha). For each subdivision, we generate a mock \ha~sample using our \wom~model and measure its corresponding SFRF. We then integrate the full SFRF to measure star-formation rate densities (SFRDs).

Figures \ref{fig:SFRD_EWs} and \ref{fig:SFRD_ssfr_mass} show our SFRDs for the full sample as a {\it light blue, dashed} line along with comparisons to other \ha~measurements and the \oii~SFRD evolution of \citet{Khostovan2015} (constrained up to $z \sim 5$). The {\it light blue circles} are SFRDs for the full \ha~population at specific redshifts corresponding to the HiZELS and DAWN samples using the \wo$(M)$ power-law fit described in Table \ref{table:EW_correlation} and Equation \ref{eqn:powerlaw_mass}. We find that our \ha~SFRDs from both \wom~and \wo($M$) models are in strong agreement with the observed \ha~SFRDs \citep{Ly2007,Westra2010,Sobral2014,Sobral2015,Stroe2015,Harish2020,Khostovan2020,VilellaRojo2021} suggesting that both approaches are tracing the observed cosmic SFRD evolution. However, we do note that the resulting SFRD at $z \sim 0.84$ and $1.47$ is underestimated in respect to the HiZELS measurement and may be linked to the underestimation in the faint-end shown in the SFRFs (Figure \ref{fig:SFRFs}). Overall, our model does predict the cosmic star formation history for the full population of \ha~emitters up to $z \sim 2.2$.

We first investigate the relative contribution of \ha~emitters based on their rest-frame EWs as shown in Figure \ref{fig:SFRD_EWs} and Table \ref{table:SFRD_EWbins}. We find that at $z > 1.5$, \ha~emitters with rest-frame EW $>200$\AA~contribute $\sim 40$ percent of the total cosmic star-formation activity. This is somewhat expected given that the \ha~EW distributions become wider (higher \wo) with increasing redshift. However, if we consider \ha~EW as a proxy for comparing instantaneous-to-past star formation activity, then this would suggest bursty, star-forming galaxies play a major role at the peak of cosmic star-formation activity. 

Furthermore, at $z > 1.5$ the main star-formation contribution is found to reside in EW $> 100$\AA~\ha~emitters ($\sim 60 - 65$ percent); however, the role of such high EW systems decreases with decreasing redshift. Between $z \sim 0.4$ and $1.5$, half of cosmic star-formation activity is found within \ha~emitters with rest-frame EW between $25$\AA~and $100$\AA. This suggests that the bulk of star formation activity shifts from extreme emission line galaxies to intermediate systems (e.g., `normal'/main-sequence galaxies) signifying a population shift.

By $z \sim 0.4$, cosmic star-formation activity is primarily subdivided between low EW systems ($0\textrm{\AA} < \textrm{EW}_0 < 25 \textrm{\AA}$) and intermediate systems ($25\textrm{\AA} < \textrm{EW}_0 < 100\textrm{\AA}$), with star formation in $z < 0.4$ SF-galaxies found in low EW systems. This would signify that the main contributor of star-formation activity at $z < 1.5$ is primarily from increasingly less active, low EW and SFR \ha~systems while at $z > 1.5$ the main contribution is from \ha~emitters exhibiting bursty, episodic SF activity producing high EWs. The {\it left} panel of Figure \ref{fig:SFRD_EWs} also shows how \ha~emitters with varying EW limits contribute to overall star-formation activity and shows the importance of high EW sources at $z > 1.5$ while low EW systems have a more significant role at lower redshifts.

Indeed if we subdivide our sample in sSFR, we also see a similar result as shown in the {\it left} panel of Figure \ref{fig:SFRD_ssfr_mass}. At $z > 1.5$, we find \ha~emitters with $10^{-8.5} < \textrm{sSFR}/\textrm{yr}^{-1} < 10^{-7.0}$ contribute half of cosmic star-formation activity. This suggests the main contribution to cosmic SFRD at $z \sim 2$ is from highly active star-forming galaxies that have mass-doubling timescales of $\sim 10 - 300$ Myr and are most likely systems with bursty, episodic star-formation histories. Between $z \sim 0.6$ and $z \sim 1.5$, this shifts to \ha~emitters with sSFR between $10^{-9.5}$ yr$^{-1}$ and $10^{-8.5}$ yr$^{-1}$, which represent `normal' star-forming galaxies with mass-doubling times of $\sim 300$ Myr to $3$ Gyr. At $z < 0.6$, $> 50$ percent of star-formation activity comes from \ha~emitters with $10^{-11.} < \textrm{sSFR}/\textrm{yr}^{-1} < 10^{-9.5}$ which represent star-forming galaxies that fall below the main sequence (e.g., galaxies in the process of quenching).

Figure \ref{fig:SFRD_ssfr_mass} also shows the relative contribution of \ha~emitters to cosmic star formation activity for varying stellar masses. At all redshifts up to $z \sim 2$, we find that massive, star-forming galaxies ($10. < \log_{10} M/\textrm{\msol} < 11.5$) have the highest contribution to cosmic SFRD. By $z\sim 0.4$ the relative contribution of $M \sim 10^{10.5 - 11}$ \msol~galaxies is 48 percent and drops to $\sim 34$ percent by $z \sim 2$. We find the relative contribution of intermediate ($10^{9.0 - 10.0}$ \msol) and high ($>10^{10}$ \msol) galaxies to decrease with increasing redshift while low-mass ($10^{8.0 - 9.0}$ \msol) galaxies increasingly contribute to overall cosmic SF activity with increasing redshift. The bulk of star-formation at $z > 1$ is found within $<10^{10}$ \msol~systems. This highlights the importance of low-mass galaxies in the high-$z$ Universe which will also tend to be the sources with elevated EW and sSFR (potential starburst galaxies). 

Overall, our \wom~model indicates that high EW/sSFR systems are the main contributors to cosmic star-formation activity at $z > 1.5$ primarily in low-mass galaxies.. This suggests low mass systems with high EW/sSFR at $z > 1.5$ (population of extreme emission line galaxies undergoing bursty, episodic star formation activity) are the main contributors to cosmic SFRD. At $z < 1.5$, we notice a population shift where intermediate EW/sSFR and massive systems (e.g., `normal' main-sequence galaxies) start to dominate cosmic SF activity. This continues to $z \sim 0.4$ where we find the final transition where the bulk of star-formation in the Universe is found in low EW/sSFR and massive \ha~emitters. This highlights key aspects of where star-formation activity primarily occurred at varying periods in cosmic time. %Further detailed investigation of the star-formation histories for different subdivisions of galaxies based on EW and sSFR are needed to understand how 

\begin{table*}
	\centering
	\caption{Star-Formation Rate Densities for varying \ha~Equivalent Width limits.  Each column in the upper table represents the SFRD assuming a specific \ewr~limit and the lower table represents the ratio in comparison to $\dot{\rho}_\textrm{All}$ which is assumed to be the SFRD at the corresponding redshift for \ewr$> 0$\AA.} 
	\label{table:SFRD_EWlimits}
	{\renewcommand{\arraystretch}{1.3}
	\begin{tabular*}{\textwidth}{@{\extracolsep{\fill}} c c c c c c }
			\hline
			$z$ & $\log_{10} \dot{\rho}_{\textrm{EW}_0 > 0\textrm{\AA}}$ & $\log_{10} \dot{\rho}_{\textrm{EW}_0 > 25\textrm{\AA}}$ & $\log_{10} \dot{\rho}_{\textrm{EW}_0 > 50\textrm{\AA}}$ & $\log_{10} \dot{\rho}_{\textrm{EW}_0 > 100\textrm{\AA}}$ & $\log_{10} \dot{\rho}_{\textrm{EW}_0 > 200\textrm{\AA}}$\\
					&  (\msol~yr$^{-1}$ Mpc$^{-3}$) &  (\msol~yr$^{-1}$ Mpc$^{-3}$) &  (\msol~yr$^{-1}$ Mpc$^{-3}$) &  (\msol~yr$^{-1}$ Mpc$^{-3}$) &  (\msol~yr$^{-1}$ Mpc$^{-3}$)\\
			\hline
			$0.40$ & $-1.94^{+0.06}_{-0.07}$ & $-2.21^{+0.10}_{-0.12}$ & $-2.59^{+0.13}_{-0.13}$ & $-3.23^{+0.13}_{-0.10}$ & $-4.23^{+0.17}_{-0.14}$ \\
			$0.62$ & $-1.72^{+0.07}_{-0.08}$ & $-1.87^{+0.10}_{-0.11}$ & $-2.12^{+0.14}_{-0.12}$ & $-2.61^{+0.21}_{-0.13}$ & $-3.41^{+0.33}_{-0.24}$ \\
			$0.84$ & $-1.55^{+0.06}_{-0.05}$ & $-1.67^{+0.08}_{-0.07}$ & $-1.84^{+0.11}_{-0.08}$ & $-2.16^{+0.17}_{-0.11}$ & $-2.65^{+0.29}_{-0.22}$ \\
			$1.47$ & $-1.22^{+0.13}_{-0.06}$ & $-1.25^{+0.14}_{-0.07}$ & $-1.31^{+0.17}_{-0.09}$ & $-1.43^{+0.21}_{-0.13}$ & $-1.64^{+0.30}_{-0.23}$ \\
			$2.23$ & $-0.90^{+0.07}_{-0.05}$ & $-0.93^{+0.07}_{-0.05}$ & $-0.97^{+0.09}_{-0.06}$ & $-1.09^{+0.11}_{-0.08}$ & $-1.31^{+0.17}_{-0.12}$ \\
	    	\hline
			$z$ & $\dot{\rho}_{> 0\textrm{\AA}}/\dot{\rho}_{\textrm{All}}$ & $\dot{\rho}_{> 25\textrm{\AA}}/\dot{\rho}_{\textrm{All}}$ & $\dot{\rho}_{> 50\textrm{\AA}}/\dot{\rho}_{\textrm{All}}$ & $\dot{\rho}_{> 100\textrm{\AA}}/\dot{\rho}_{\textrm{All}}$ & $\dot{\rho}_{> 200\textrm{\AA}}/\dot{\rho}_{\textrm{All}}$\\
			&   (\%) &   (\%) &   (\%) &  (\%)  &  (\%)\\				
			\hline
			$0.40$ & $100$ & $53.3^{+5.6}_{-6.2}$ & $22.5^{+4.0}_{-2.9}$ & $5.2^{+1.2}_{-0.6}$ & $0.5^{+0.2}_{-0.1}$ \\
			$0.62$ & $100$ & $70.5^{+5.7}_{-5.8}$ & $39.8^{+7.5}_{-4.7}$ & $12.9^{+5.7}_{-2.4}$ & $2.0^{+2.3}_{-0.9}$ \\
			$0.84$ & $100$ & $76.9^{+4.0}_{-3.3}$ & $51.4^{+6.9}_{-4.1}$ & $24.6^{+8.3}_{-4.3}$ & $8.0^{+6.8}_{-3.0}$ \\
			$1.47$ & $100$ & $92.8^{+2.2}_{-1.4}$ & $80.9^{+6.0}_{-3.7}$ & $60.7^{+12.4}_{-8.3}$ & $37.5^{+18.7}_{-12.2}$ \\
			$2.23$ & $100$ & $95.0^{+1.1}_{-1.1}$ & $85.3^{+3.0}_{-2.4}$ & $65.7^{+6.3}_{-4.1}$ & $39.4^{+10.4}_{-6.7}$ \\
			\hline
		\end{tabular*}    
	}
\end{table*}

\begin{table*}
	\centering
	\caption{Star-Formation Rate Densities for varying \ha~Equivalent Width ranges.  For each \ewr~bin we include the SFRD and the ratio with $\dot{\rho}_\textrm{All}$ which is defined as the SFRD with \ewr$>0$\AA~shown in Table \ref{table:SFRD_EWlimits}.}
	\label{table:SFRD_EWbins}
	{\renewcommand{\arraystretch}{1.3}
	\begin{tabular*}{\textwidth}{@{\extracolsep{\fill}} c c c c c c c c c}
		\hline
		& \multicolumn{2}{c}{$0 < \textrm{EW}_0/\textrm{\AA} < 25$} & \multicolumn{2}{c}{$25 < \textrm{EW}_0/\textrm{\AA} < 100$} & \multicolumn{2}{c}{$100 < \textrm{EW}_0/\textrm{\AA} < 200$} & \multicolumn{2}{c}{$200 < \log_{10} \textrm{EW}_0/\textrm{\AA}$} \\
		$z$ & $\log_{10} \dot{\rho}$ &  $\dot{\rho}/\dot{\rho}_{\textrm{All}}$ & $\log_{10} \dot{\rho}$ &  $\dot{\rho}/\dot{\rho}_{\textrm{All}}$ & $\log_{10} \dot{\rho}$ &  $\dot{\rho}/\dot{\rho}_{\textrm{All}}$ & $\log_{10} \dot{\rho}$ &  $\dot{\rho}/\dot{\rho}_{\textrm{All}}$ \\
		&  (\msol~yr$^{-1}$ Mpc$^{-3}$) &  (\%) &  (\msol~yr$^{-1}$ Mpc$^{-3}$) &   (\%)  &  (\msol~yr$^{-1}$ Mpc$^{-3}$) &   (\%) & (\msol~yr$^{-1}$ Mpc$^{-3}$) &   (\%)\\
		\hline				
		$0.40$ & $-2.27^{+0.01}_{-0.03}$ & $46.8^{+1.4}_{-3.1}$ & $-2.27^{+0.11}_{-0.12}$ & $47.1^{+12.3}_{-13.0}$ & $-3.27^{+0.13}_{-0.11}$ & $4.7^{+1.4}_{-1.2}$ & $-4.23^{+0.17}_{-0.14}$  & $0.5^{+0.2}_{-0.1}$ \\
		$0.62$ & $-2.25^{+0.01}_{-0.04}$ & $29.5^{+0.8}_{-2.8}$ & $-1.96^{+0.09}_{-0.14}$ & $57.6^{+11.0}_{-18.5}$ & $-2.68^{+0.17}_{-0.13}$ & $10.9^{+4.5}_{-3.4}$ & $-3.41^{+0.33}_{-0.24}$ & $2.0^{+2.3}_{-0.9}$ \\
		$0.84$ & $-2.19^{+0.02}_{-0.03}$ & $23.1^{+0.9}_{-1.7}$ & $-1.84^{+0.07}_{-0.10}$ & $52.3^{+7.4}_{-11.9}$ & $-2.33^{+0.11}_{-0.10}$ & $16.6^{+4.1}_{-3.9}$ & $-2.65^{+0.29}_{-0.22}$ & $8.0^{+6.8}_{-3.0}$ \\
		$1.47$ & $-2.36^{+0.03}_{-0.05}$ & $7.2^{+0.4}_{-0.9}$ & $-1.71^{+0.05}_{-0.07}$ & $31.9^{+2.6}_{-5.3}$ & $-1.86^{+0.07}_{-0.09}$ & $23.0^{+3.1}_{-4.8}$ & $-1.64^{+0.30}_{-0.23}$  & $37.5^{+18.7}_{-12.2}$ \\
		$2.23$ & $-2.21^{+0.06}_{-0.05}$ & $4.9^{+0.6}_{-0.6}$ & $-1.44^{+0.02}_{-0.05}$ & $29.4^{+1.2}_{-3.5}$ & $-1.48^{+0.07}_{-0.10}$ & $26.3^{+3.9}_{-5.9}$ & $-1.31^{+0.17}_{-0.12}$ & $39.4^{+10.4}_{-6.7}$ \\
		\hline		
	\end{tabular*}    
	}
\end{table*}

\begin{table*}
	\centering
	\caption{Star-Formation Rate Densities for varying Stellar Masses. Same format as in Table \ref{table:SFRD_EWbins}.}
	\label{table:SFRD_stellar_mass}	
	{\renewcommand{\arraystretch}{1.3}
	\begin{tabular*}{\textwidth}{@{\extracolsep{\fill}} c c c c c c c}
		\hline
		& \multicolumn{2}{c}{$8.0 < \log_{10} \textrm{M}/\textrm{\msol} < 9.0$} & \multicolumn{2}{c}{$9.0 < \log_{10} \textrm{M}/\textrm{\msol} < 10.0$} & \multicolumn{2}{c}{$10.0 < \log_{10} \textrm{M}/\textrm{\msol} < 11.5$} \\
		$z$ & $\log_{10} \dot{\rho}$ &  $\dot{\rho}/\dot{\rho}_{\textrm{All}}$ & $\log_{10} \dot{\rho}$ &  $\dot{\rho}/\dot{\rho}_{\textrm{All}}$ & $\log_{10} \dot{\rho}$ &  $\dot{\rho}/\dot{\rho}_{\textrm{All}}$ \\
		&  (\msol~yr$^{-1}$ Mpc$^{-3}$) &  (\%) &  (\msol~yr$^{-1}$ Mpc$^{-3}$) &   (\%)  &  (\msol~yr$^{-1}$ Mpc$^{-3}$) &   (\%) \\
		\hline		
		$0.40$ & $-2.78^{+0.01}_{-0.03}$ & $14.5^{+0.4}_{-1.0}$ &$-2.49^{+0.11}_{-0.12}$ & $28.2^{+7.4}_{-7.8}$ &$-2.25^{+0.13}_{-0.11}$ & $48.6^{+14.7}_{-12.4}$\\
		$0.62$ & $-2.55^{+0.01}_{-0.04}$ & $14.6^{+0.4}_{-1.4}$ &$-2.27^{+0.09}_{-0.14}$ & $28.0^{+5.3}_{-9.0}$ &$-2.03^{+0.17}_{-0.13}$ & $48.6^{+20.1}_{-15.2}$\\
		$0.84$ & $-2.31^{+0.02}_{-0.03}$ & $17.5^{+0.7}_{-1.3}$ &$-2.11^{+0.07}_{-0.10}$ & $27.8^{+3.9}_{-6.3}$ &$-1.93^{+0.11}_{-0.10}$ & $42.6^{+10.4}_{-10.1}$\\
		$1.47$ & $-1.86^{+0.03}_{-0.05}$ & $22.9^{+1.4}_{-2.7}$ &$-1.78^{+0.05}_{-0.07}$ & $27.4^{+2.2}_{-4.6}$ &$-1.75^{+0.07}_{-0.09}$ & $29.6^{+4.0}_{-6.1}$\\
		$2.23$ & $-1.58^{+0.06}_{-0.05}$ & $21.0^{+2.6}_{-2.6}$ &$-1.49^{+0.02}_{-0.05}$ & $25.9^{+1.1}_{-3.1}$ &$-1.37^{+0.07}_{-0.10}$ & $34.3^{+5.1}_{-7.7}$\\
		\hline
	\end{tabular*}    
	}
\end{table*}

\begin{table*}
	\centering
	\caption{Star-Formation Rate Densities for varying Specific Star Formation Rates. Same format as in Table \ref{table:SFRD_EWbins}.} 
	\label{table:SFRD_sSFR}		
	{\renewcommand{\arraystretch}{1.3}
	\begin{tabular*}{\textwidth}{@{\extracolsep{\fill}} c c c c c c c}
		\hline
			 	& \multicolumn{2}{c}{$-11. < \log_{10} \textrm{sSFR}/\textrm{yr}^{-1} < -9.5$} & \multicolumn{2}{c}{$-9.5 < \log_{10} \textrm{sSFR}/\textrm{yr}^{-1} < -8.5$} & \multicolumn{2}{c}{$-8.5 < \log_{10} \textrm{sSFR}/\textrm{yr}^{-1} < -7.0$} \\
		$z$ & $\log_{10} \dot{\rho}$ &  $\dot{\rho}/\dot{\rho}_{\textrm{All}}$ & $\log_{10} \dot{\rho}$ &  $\dot{\rho}/\dot{\rho}_{\textrm{All}}$ & $\log_{10} \dot{\rho}$ &  $\dot{\rho}/\dot{\rho}_{\textrm{All}}$ \\
		&  (\msol~yr$^{-1}$ Mpc$^{-3}$) &  (\%) &  (\msol~yr$^{-1}$ Mpc$^{-3}$) &   (\%)  &  (\msol~yr$^{-1}$ Mpc$^{-3}$) &   (\%) \\
		\hline
		$0.40$ & $-2.14^{+0.01}_{-0.03}$ & $63.5^{+1.9}_{-4.2}$ &$-2.41^{+0.11}_{-0.12}$ & $33.8^{+8.8}_{-9.3}$ &$-3.86^{+0.13}_{-0.11}$ & $1.2^{+0.4}_{-0.3}$\\
		$0.62$ & $-2.05^{+0.01}_{-0.04}$ & $47.0^{+1.2}_{-4.4}$ &$-2.03^{+0.09}_{-0.14}$ & $48.9^{+9.3}_{-15.7}$ &$-3.15^{+0.17}_{-0.13}$ & $3.7^{+1.5}_{-1.2}$\\
		$0.84$ & $-2.03^{+0.02}_{-0.03}$ & $33.7^{+1.3}_{-2.5}$ &$-1.83^{+0.07}_{-0.10}$ & $53.2^{+7.5}_{-12.1}$ &$-2.43^{+0.11}_{-0.10}$ & $13.2^{+3.2}_{-3.1}$\\
		$1.47$ & $-2.26^{+0.03}_{-0.05}$ & $9.1^{+0.6}_{-1.1}$ &$-1.60^{+0.05}_{-0.07}$ & $41.5^{+3.3}_{-6.9}$ &$-1.54^{+0.07}_{-0.09}$ & $47.9^{+6.5}_{-9.9}$\\
		$2.23$ & $-2.14^{+0.06}_{-0.05}$ & $5.8^{+0.7}_{-0.7}$ &$-1.36^{+0.02}_{-0.05}$ & $35.2^{+1.4}_{-4.2}$ &$-1.17^{+0.07}_{-0.10}$ & $54.3^{+8.1}_{-12.2}$\\
		\hline
	\end{tabular*}    
	}
\end{table*}

\subsection{sSFR Evolution: The role of selection bias}

\begin{figure}
	\centering
	\includegraphics[width=\columnwidth]{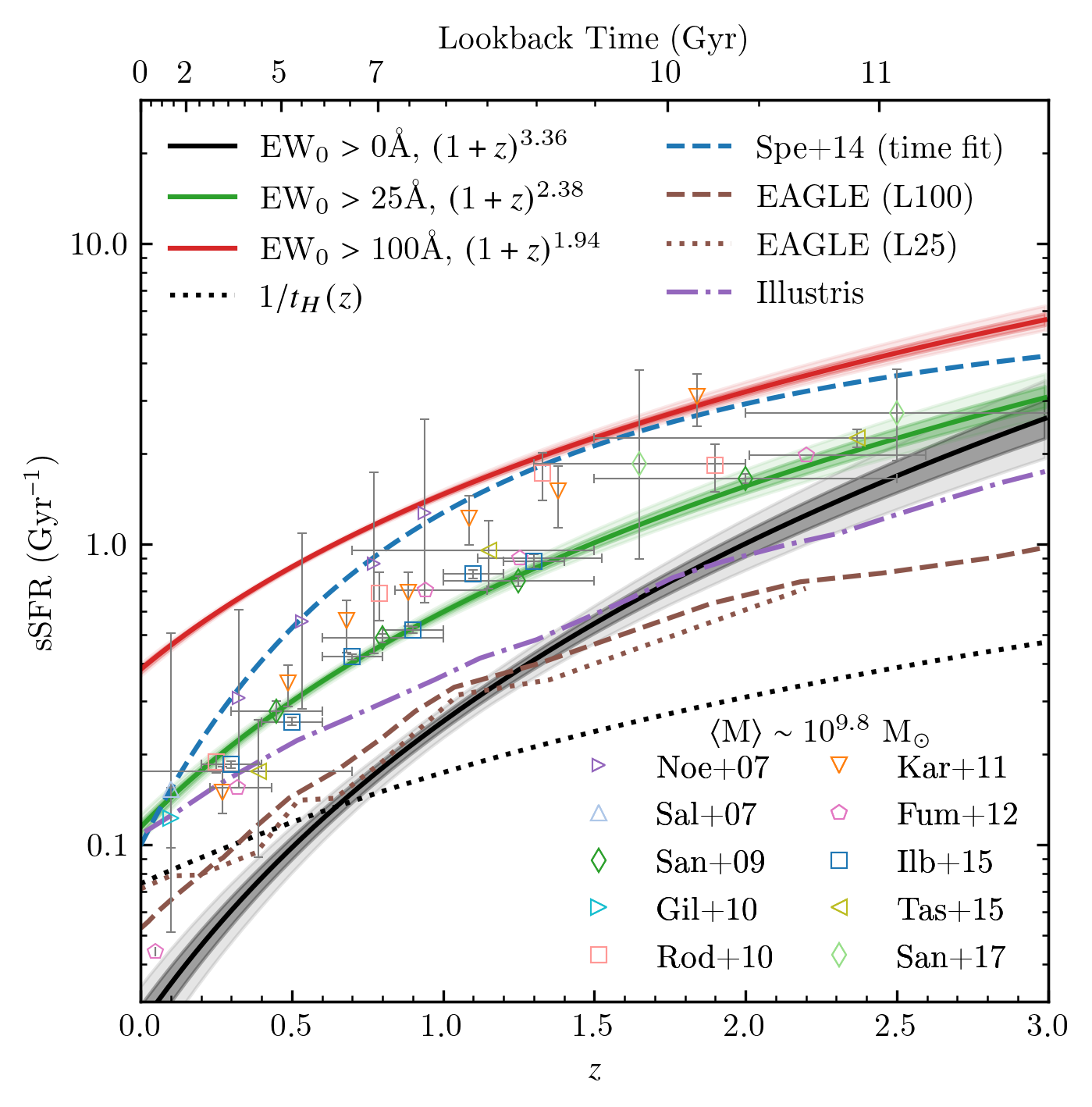}
	\caption{Cosmic sSFR evolution for star-forming galaxies between $10^{9.5 - 10}$ \msol~with predictions from our \wom~model. Included are observation measurements (e.g., \citealt{Noeske2007,Salim2007,Santini2009,Gilbank2010,Rodighiero2010,Karim2011,Fumagalli2012,Ilbert2015,Tasca2015,Santini2017}), the empirically constrained main sequence model of \citet{Speagle2014}, and predictions from cosmological hydrodynamical simulations EAGLE \citep{Furlong2015} and Illustris \citep{Sparre2015}. We find our best-fit \wom~model for the full range of \ha~emitters (\ewr$>0$\AA; {\it black line}) is consistent with simulations and a $(1+z)^{3.36}$ evolution. Our $z = 0$ prediction is also similar to \citet{Fumagalli2012} which had a \ewr$>3$\AA~limit. However, applying varying \ewr~cuts raises the normalization and decreases the slope. Overall, this shows how we can resolve the tension between observations and simulations in the typical sSFR cosmic evolution by applying selection limits to an intrinsic sample of star-forming galaxies.}
	\label{fig:ssfr_bias}
\end{figure}

Observationally, the typical sSFR at a stellar mass of $\sim 10^{9.5 - 10}$ \msol~is found to evolve as $(1+z)^{2.4 - 3.0}$ at $z < 2.2$ \citep{Ilbert2015,Faisst2016} and $(1+z)^{1.1 - 1.5}$ up to $z \sim 7$ \citep{Tasca2015,Faisst2016,Davidzon2018}. Figure \ref{fig:ssfr_bias} shows the cosmic sSFR evolution where observations find more than an order-of-magnitude increase from the local Universe to cosmic noon. 

Large hydronamical simulations such as EAGLE and Illustris \citep{Furlong2015,Sparre2015} are able to predict the shape of this redshift evolution. However, there is a factor of $\sim 2 - 3$ discrepancy in the typical sSFR (normalization of the trend) between observations and simulations. We explore the cause of this discrepancy by using our observationally-constrained \wom~model to investigate the cosmic evolution of sSFR and how selection biases could affect the underlying measurement of typical sSFR at fixed stellar mass. 

We measure sSFR using the assigned stellar mass for each \ha~emitter and its corresponding \ha~SFR which is determined by converting stellar mass to \ha~luminosity via the assigned EW and applying the \citet{Kennicutt1998} calibration assuming Chabrier IMF. The typical sSFR at a given redshift and stellar mass range is determined by taking the median sSFR of the mock sample for that respective redshift and stellar mass range. We find that the typical sSFR from our \wom~model can be parameterized as:
\begin{eqnarray}
\textrm{sSFR}(z,M) &=& \textrm{sSFR}_{10} \bigg(\frac{M}{10^{10} \textrm{\msol}}\bigg)^{\beta_0} (1+z)^{\xi(M)} \nonumber\\
\xi(M) &=& \xi_{10} + \beta\log_{10}\bigg(\frac{M}{10^{10} \textrm{\msol}}\bigg)
\label{eqn:ssfr_evol}
\end{eqnarray}
where sSFR$_{10}$ is the typical sSFR at $z = 0$ and stellar mass of $10^{10}$ \msol, $\beta_0$ is the sSFR -- stellar mass correlation slope at $z = 0$, $\xi_{10}$ is the redshift-dependent sSFR slope for $10^{10}$ \msol~star-forming galaxies, and $\beta$ defines how the redshift evolution slope changes with stellar mass. Essentially, Equation \ref{eqn:ssfr_evol} is a combination of the sSFR -- stellar mass correlation and cosmic sSFR evolution similar to how we model the \wo-- stellar mass correlation and redshift evolution. 

Figure \ref{fig:ssfr_bias} shows the typical sSFR within the range of $10^{9.5 - 10}$ \msol~ with a median stellar mass of $\sim 10^{9.8}$ \msol~for three limiting rest-frame EW limits: 0\AA, 25\AA, and 100\AA. In all three cases we do not consider any \ha~limiting line flux; however, we do include the cosmic sSFR evolution parameterization in Table \ref{table:ssfr_params} for all cases including EW and \ha~line fluxes limits. The `intrinsic' sSFR evolution is described as the case where EW$_0 > 0$\AA~and no \ha~line flux limits are applied which we show as a {\it black solid} line in Figure \ref{fig:ssfr_bias}. We find that our \wom~model shows an increasing sSFR with increasing redshift where we find close to a two order-of-magnitude increase by $z \sim 2.2$ and described by a $(1+z)^{3.36}$ evolution. This is somewhat steeper compared to what is found in observations \citep{Ilbert2015,Faisst2016}.

We include observational measurements of the typical sSFR within the same stellar mass range of our measurements with a comparable median stellar mass of $\langle M \rangle \sim 10^{9.8}$ \msol~in Figure \ref{fig:ssfr_bias}. \citep{Noeske2007,Salim2007,Santini2009,Gilbank2010,Rodighiero2010,Karim2011,Fumagalli2012,Ilbert2015,Tasca2015,Santini2017}. We find that our `intrinsic' sSFR evolution measurement is consistently below observations by a factor of $\sim 2 - 3$.  The only exception is the $z \sim 0$ measurement of \citet{Fumagalli2012} which has a limiting \ha~EW$_0 > 3$\AA~ and is also a factor of $\sim 3$ below the $z \sim 0$ \citet{Salim2007} and \citet{Gilbank2010} measurements. Comparing to simulations, we find that at $z > 1.5$ we are in agreement with Illustris \citep{Sparre2015} while at $z < 1.5$ we are in agreement with the L25 and L100 simulations from EAGLE \citep{Furlong2015}. %{\color{red} WHY?? IT PROBABLY IS DUE TO DIFFERING PHYSICS? ZZZ}. 

Overall, our `intrinsic' model better matches simulations while the disagreement with observations is due to selection bias effects in the latter. For example, the $z \sim 0$ \citet{Fumagalli2012} measurement is in agreement with our `intrinsic' sSFR evolution with the reason being that embedded in this measurement is a low limiting $EW_0 > 3$\AA~limit which is close to the $EW_0 > 0$\AA~assumed in our `intrinsic' model. \citet{Fumagalli2012} is also not in agreement with the $z \sim 0$ measurements of \citet{Salim2007} and \citet{Gilbank2010}. We also see a scatter of $\sim 0.3$ dex in the literature measurements which is primarily arising from different sample selections and SF calibrations used. 

Figure \ref{fig:ssfr_bias} also shows the cosmic sSFR evolution for the case of a $>25$\AA~and $>100$\AA~limiting rest-frame EW case as a {\it solid green} and {\it red} line, respectively. We find that the $> 25$\AA~case is in agreement with observations and has a $(1+z)^{2.38}$ evolution consistent with past studies \citep{Ilbert2015,Faisst2016}. This highlights how applying a selection bias to our mock \ha~samples can reconcile the discrepancy between our `intrinsic' sSFR evolution (in a agreement with simulations) and observations. It also further suggests how selection biases can play an important role in how we measure the cosmic sSFR evolution.

Interestingly, applying an $EW_0 > 100$\AA~cut results in a $(1+z)^{1.94}$ evolution somewhat consistent with the $(1+z)^{1.1 - 1.5}$ evolution measured at $z > 2.2$  \citep{Tasca2015,Faisst2016,Davidzon2018}. Although our models are observationally-constrained up to $z = 2.23$, it is interesting to note that the \ha$+$\nii~EWs measured via {\it Spitzer} IRAC color excess measured at $z > 2$ are primarily biased towards high EWs such that the shallower  $(1+z)^{1.1  - 1.5}$ evolution is probably due to changing selection limits with increasing redshift. Overall, we find that the major culprit in the discrepancy between observations and simulations when it comes to the cosmic sSFR evolution is the result of selection limits causing elevated median sSFR.

\begin{table*}
	\centering
	\caption{Parameters describing the Cosmic Specific Star Formation Rate Evolution in Equation \ref{eqn:ssfr_evol}. These measurements are based on our observationally-constrained redshift and stellar mass-dependent \ha~\wom~model. The `intrinsic' sSFR(M,$z$) model is shown in the first row where the rest-frame \ha~EW limit is set to $0$\AA~and no \ha~flux limits are applied. All other measurements below describe how selection-biases in the form of \ha~EW$_0$ and line flux cuts can affect our view of the cosmic sSFR evolution. The models are normalized to the sSFR for a $10^{10}$\msol~\ha~emitter at $z = 0$, sSFR$_{10}$. The sSFR--stellar mass anti-correlation at $z = 0$ is defined by the slope, $\beta_0$ and $\beta$ describes the sSFR redshift evolution at varying stellar masses normalized to the case of a $10^{10}$ \msol~\ha~emitter, $\xi_{10}$.}
	\label{table:ssfr_params}
	{\renewcommand{\arraystretch}{1.3}
	\begin{tabular*}{\textwidth}{@{\extracolsep{\fill}} c c c c c c}
		\hline
		\ha~EW$_0$ Limit & Obs \ha~Flux Limit & sSFR$_{10}$ & $\xi_{10}$ & $\beta_0$ & $\beta$ \\
		(\AA) & (erg s$^{-1}$ cm$^{-2}$) & (Gyr$^{-1}$)	& & & \\
		\hline
		$0$ & None & $0.022^{+0.004}_{-0.004}$ & $3.21^{+0.22}_{-0.22}$ & $-0.25^{+0.04}_{-0.04}$ & $-0.59^{+0.16}_{-0.18}$ \\
		& $1 \times 10^{-17}$ & $0.023^{+0.004}_{-0.003}$ & $3.63^{+0.19}_{-0.18}$ & $-0.30^{+0.03}_{-0.04}$ & $-0.73^{+0.12}_{-0.12}$ \\
		& $3 \times 10^{-17}$ & $0.025^{+0.004}_{-0.003}$ & $4.10^{+0.16}_{-0.15}$ & $-0.36^{+0.03}_{-0.03}$ & $-0.73^{+0.10}_{-0.10}$ \\
		& $1 \times 10^{-16}$ & $0.030^{+0.003}_{-0.004}$ & $4.71^{+0.14}_{-0.12}$ & $-0.43^{+0.03}_{-0.03}$ & $-0.66^{+0.10}_{-0.07}$ \\
		\hline
		$25$ & None & $0.106^{+0.005}_{-0.006}$ & $2.22^{+0.09}_{-0.08}$ & $-0.13^{+0.01}_{-0.02}$ & $-0.65^{+0.08}_{-0.10}$ \\
		& $1 \times 10^{-17}$ & $0.102^{+0.005}_{-0.005}$ & $2.39^{+0.08}_{-0.08}$ & $-0.12^{+0.01}_{-0.01}$ & $-0.92^{+0.06}_{-0.07}$ \\
		& $3 \times 10^{-17}$ & $0.097^{+0.004}_{-0.004}$ & $2.80^{+0.07}_{-0.07}$ & $-0.15^{+0.01}_{-0.01}$ & $-1.11^{+0.04}_{-0.07}$ \\
		& $1 \times 10^{-16}$ & $0.092^{+0.003}_{-0.004}$ & $3.56^{+0.06}_{-0.04}$ & $-0.25^{+0.02}_{-0.01}$ & $-1.03^{+0.04}_{-0.07}$ \\
		\hline
		$100$ & None & $0.358^{+0.008}_{-0.008}$ & $1.80^{+0.04}_{-0.04}$ & $-0.12^{+0.01}_{-0.01}$ & $-0.54^{+0.05}_{-0.05}$ \\
		& $1 \times 10^{-17}$ & $0.353^{+0.008}_{-0.008}$ & $1.85^{+0.04}_{-0.04}$ & $-0.11^{+0.01}_{-0.01}$ & $-0.64^{+0.04}_{-0.05}$ \\
		& $3 \times 10^{-17}$ & $0.339^{+0.008}_{-0.007}$ & $2.04^{+0.04}_{-0.04}$ & $-0.09^{+0.01}_{-0.01}$ & $-0.92^{+0.04}_{-0.04}$ \\
		& $1 \times 10^{-16}$ & $0.311^{+0.006}_{-0.006}$ & $2.53^{+0.03}_{-0.03}$ & $-0.11^{+0.01}_{-0.01}$ & $-1.18^{+0.05}_{-0.04}$ \\							
		\hline
	\end{tabular*}
	}
\end{table*}

\section{Forecasting Next-Generation Surveys: \textit{Roman} and  \textit{Euclid}}
\label{sec:forecast}
Future slitless grism surveys using {\it Nancy Grace Roman Space Telescope} and {\it Euclid} will be able to observe sizeable samples of star-forming galaxies that would enhance our understanding of galaxy evolution physics. Accurate estimation of the number counts for planned surveys using these state-of-the-art facilities is crucial in regards to survey designing and planning of science goals and objectives. A simple, back-of-the-envelope estimation of the expected number counts and redshift distribution can be determined by using reported line luminosity functions in combination with a limiting line flux threshold. However, slitless grism surveys have a resolution limit that can be interpreted as a minimum limiting EW threshold. 

For example, {\it HST}/WFC3 G141 has a resolving power of $R \sim 130$ for point-sources which acts also as an observed EW limiting threshold of 85\AA~to 130\AA~at $1.1\mu$m and $1.7\mu$m, respectively. For $z \sim 0.7$ -- 1.6 \ha~emitters, this corresponds to a rest-frame EW threshold of $\sim 50$\AA. Using our best-fit \wom~model, we find the typical EW ranges from $\sim 23$\AA~to $\sim 90$\AA~for 10$^{10}$ \msol~\ha~emitters from $z \sim 0.7$ to $1.6$, respectively. Given that \wo~corresponds to 50 percent of the total exponential EW distributions, it implies that a {\it HST}/WFC3 G141 survey of \ha~emitters could miss 88 and 43 percent of the total \ha~EW distribution at $z \sim 0.7$ and $1.6$, respectively. The low-$z$ end is especially affected given that EW distributions become narrower (lower \wo) with decreasing redshift such that the $50$\AA~rest-frame EW limit becomes an important factor towards low-$z$. 

This effect of limiting EW weakens with decreasing stellar mass. For example, a similar grism survey that observes $10^8$ \msol~\ha~emitters will miss 27 percent and 14 percent of the total \ha~EW distribution at $z \sim 0.7$ and $1.66$, respectively, which is due to wider distributions (higher \wo) at lower stellar masses. This highlights the importance of incorporating EW distributions in survey number count forecasting for grism surveys. 

In this section, we present our predicted number counts and EW properties for large grism surveys planned with {\it Nancy Grace Roman Space Telescope} and {\it Euclid}. We use our best-fit \wom~model to generate mock samples of \ha~emitters corresponding to the redshift range associated with the planned grism surveys. We also convolve the grism throughputs with the intrinsic (not corrected for dust) \ha$+$\nii~luminosities in our mock samples to measure the expected \ha$+$\nii~line fluxes that would be observed in such surveys. For each telescope, we consider three different \ha$+$\nii~line flux limits: $>2 \times 10^{-16}$ erg s$^{-1}$ cm$^{-2}$ (``shallow''), $>1 \times 10^{-16}$ erg s$^{-1}$ cm$^{-2}$ (``intermediate''), and $>5 \times 10^{-17}$ erg s$^{-1}$ cm$^{-2}$ (``deep''). The limiting observed EW thresholds are also included to take into account missing low EW sources due to the changing EW distributions.

\begin{figure*}
	\centering
	\includegraphics[width=\textwidth]{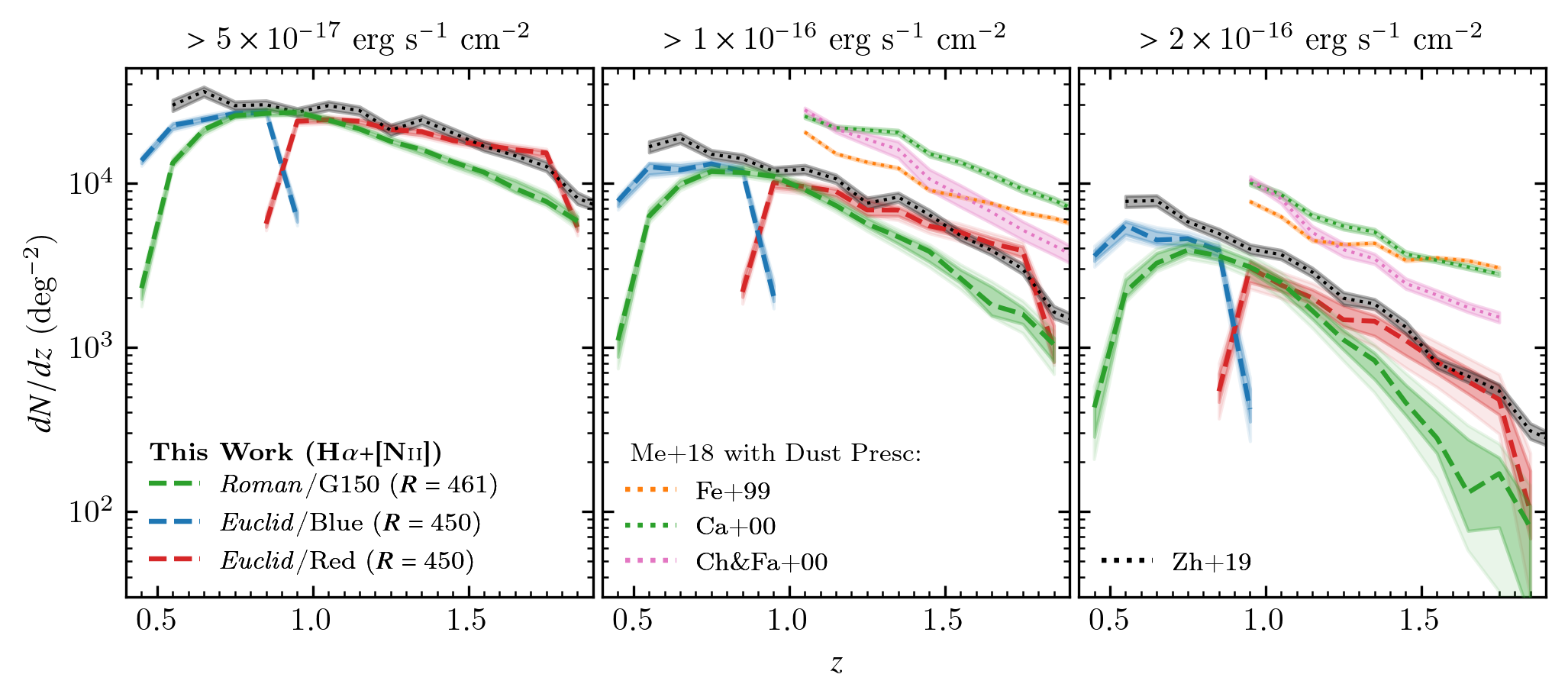}
	\caption{Predicted redshift distributions for {\it Roman} and {\it Euclid} based on our empirical intrinsic \ha$+$\nii~\wo($M,z$) model for three different line flux limits: $5\times10^{-17}$ (\textit{left}), $1\times10^{-16}$ (\textit{middle}), and $2\times10^{-16}$ \cgsline~(\textit{right}). The predicted distributions take into account line flux selection limits, grism throughputs, and a limiting observed EW threshold based on the resolving power, $R$.  The limiting EW effect is most pronounced at low-$z$ (\wo~decreases with decreasing redshift). This signifies the importance of taking into account such limits in number counts for planned grism surveys. {\sc GALACTICUS}-based number counts (\citealt{Merson2018,Zhai2019}; the former includes various dust prescriptions) are also found to overestimate the number of galaxies, which is most likely attributed to a combination of overestimated \ha~LFs in comparison to observations, different applied dust prescriptions, and not accounting for observational efficiency and limiting resolution. However, the main consensus is that both {\it Euclid} and {\it Roman} will be able to observe large number of $0.5 < z < 2$ \ha~emitters in a wide survey.}
	\label{fig:zdistrib_surveys}
\end{figure*}

\begin{figure*}
	\centering
	\includegraphics[width=\textwidth]{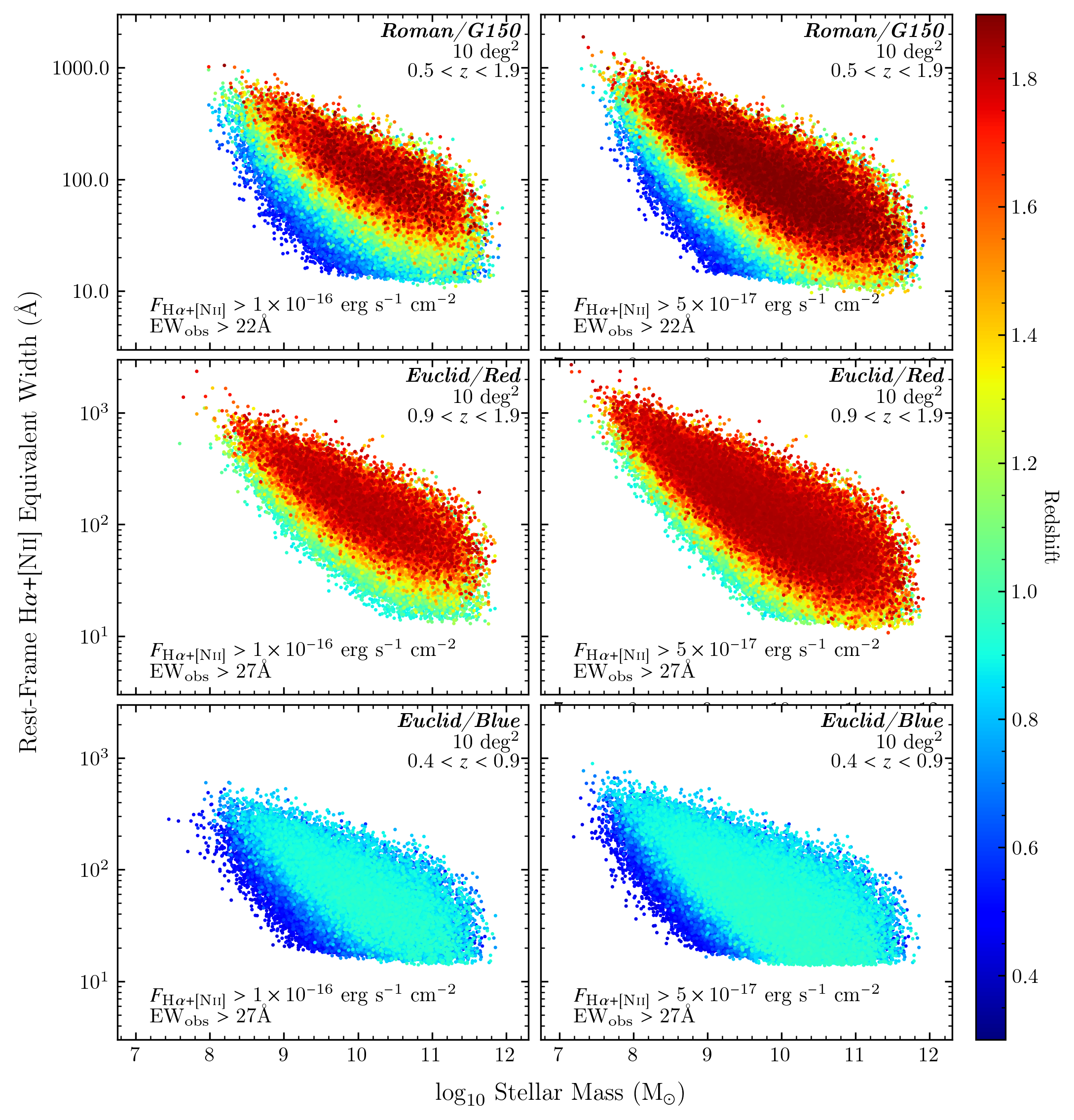}
	\caption{A mock 10 deg$^2$ {\it Roman} ({\it top}), {\it Euclid}/Red ({\it middle}), and {\it Euclid}/Blue ({\it bottom}) grism survey with the range of rest-frame \ha$+$\nii~equivalent widths, stellar mass, and redshift per survey using our \wom~model along with the grism throughput of each respective instrument. The {\it left} panels assume a $10^{-16}$ erg s$^{-1}$ cm$^{-2}$ observed \ha$+$\nii~flux limit consistent with {\it Roman } HLSS and 50\% deeper than {\it Euclid}-Wide along with a limiting EW threshold based on the limiting resolution, $R$, of the respective grism used. The {\it right} panels shows the same except for an observed flux limit of $5\times10^{-17}$ erg s$^{-1}$ cm$^{-2}$. We find both surveys cover a wide range of stellar mass and EW. {\it Euclid} is expected to find more $>1000$\AA~low-mass \ha~emitters given its increased sensitivity at high-$z$ compared to {\it Roman} at the same line flux limits. However, {\it Roman} is also sensitive to $z< 0.8$ low-mass systems with high EWs. Furthermore, {\it Euclid}-Wide has a brighter flux limit which will result in fewer $>1000$\AA~emitters compared to {\it Roman} HLSS. We note this does not take into account high EW outliers (sources that fall outside the exponential EW distribution) such that we can expect to also have high mass, high EW systems in such surveys, especially in the case where wide areas are considered.}
	\label{fig:EWdistrib_surveys}
\end{figure*}

\subsection{Nancy Grace Roman Space Telescope}

The \textit{ Nancy Grace Roman Space Telescope} is a 2.4-meter space-based observatory planned for launch in mid-2027 \citep{Akeson2019}. The primary extragalactic astrophysics instrument on-board will be its Wide Field Instrument (WFI) which will cover $0.28$ deg$^2$ per exposure and a wavelength coverage from 4800\AA~to 23000\AA~with 8 broadband imaging filters. {\it Roman}/WFI will also include a single grism, G150, centered at $1.465$\micron~and covering from 1\micron~to $1.93$\micron~with a limiting resolution of $R\sim 461$ at $1$\micron~(2 pix). The G150 wavelength coverage corresponds to $0.5 < z < 1.9$ \ha~emitters; however, the limiting resolution sets a minimum observed \ha$+$\nii~equivalent width of $\sim 22$\AA. This corresponds to a rest-frame EW limit of $\sim 15$\AA~and $\sim 8$\AA~ at $z \sim 0.5$ and $1.9$, respectively. Given that \wo~at $z \sim 0.5$ is $\sim 22$\AA~for 10$^{10}$ \msol~\ha~emitters, the expected number counts can be significantly reduced in the low-$z$, high-mass end in comparison to relying solely on past reported luminosity functions to forecast number counts.

Figure \ref{fig:zdistrib_surveys} and Table \ref{table:zdistrib_NGRST} show our expected redshift distributions for a {\it Roman}/G150 survey using our \wom~model and applying selection limits based on an observed \ha$+$\nii~emission line flux cut and the limiting resolution/EW threshold. We find our approach predicts a redshift distribution peaked at $z \sim 0.7 - 0.8$ for our shallow survey and increases to $z \sim 0.9 - 1.0$ with the deep survey. In comparison to the simulation of \citet{Zhai2019}, we are in agreement at the peak of the redshift distirbution. The discrepancy at other redshifts is primarily due to incorporating the grism throughput and limiting EW thresholds, which is evident in the drop seen at $z \sim 0.6$ and $z > 1$. We also compare to the \citet{Merson2018} simulation results which assumes three different dust prescriptions \citep{Ferrara1999,Calzetti2000,Charlot2000} used to uncorrect their intrinsic sample. All three predict higher number counts in comparison to our results and \citet{Zhai2019}. This is primarily due to the simulated \citet{Merson2018} LFs overestimating the bright-end compared to observations by almost an order-of-magnitude (depending on the assumed dust prescription; see their Figure 9). \citet{Zhai2019} also used simulated LFs that are more consistent with observations although there are redshift ranges for which the bright-end is overestimated primarily towards low-$z$ (see their Figure 1). However, the main discrepancy arises from our inclusion of limiting EW thresholds and the G150 throughput which are not included in the \citet{Zhai2019} measurements.

For our shallow survey, we find a significant drop in the number densities with increasing redshift which is due to (1) bright ELGs are fainter at high-$z$ compared to low-$z$ and (2) decreasing grism sensitivity at redder wavelengths (increasing redshift). As shown in Table \ref{table:zdistrib_NGRST}, we predict that {\it Roman}/G150 observations will observe $\sim 1600$ and $\sim 720$ \ha$+$\nii~emitters per deg$^{-2}$ at $0.5 < z < 1$ and $1 < z < 1.9$, respectively, for a bright survey. In comparison, we predict a deep survey will observe $\sim 11300$ and $\sim 12700$ \ha$+$\nii~emitters per deg$^{-2}$ at $0.5 < z < 1$ and $1 < z < 1.9$, respectively, highlighting how decreasing the line flux limit would help in observing larger numbers of high-$z$ emitters. A shallow, wide survey would need to observe an area 10 times larger than a small, deep survey to achieve comparable number counts albeit for a brighter population of \ha~emitters.

Figure \ref{fig:EWdistrib_surveys} shows the redshift distribution and range in rest-frame \ha$+$\nii~EW and stellar mass that would be observed in a 10 deg$^2$ {\it Roman} survey for the case of an intermediate and deep survey. An observed EW$ > 22$\AA~cut is applied consistent with the G150 limiting resolution. We find that the intermediate survey probes down to $\sim 10^8$ \msol~and a maximum \ha$+$\nii~EW$_0 \sim 1000$\AA~while the deep survey pushes the stellar mass limit down to $\sim 10^{7.2}$ \msol~with a maximum  \ha$+$\nii~EW$_0 \sim 2000$\AA. The line flux limit selection effect is more prevalent in the intermediate survey as it removes low EW systems above the limiting EW threshold at fixed stellar mass with increasing redshift in the high-mass end ($> 10^{10}$ \msol). By $z> 1.5$ the minimum EW$_0$ is $\sim 30$\AA~for $10^{11}$ \msol~emitters corresponding to an observed EW $\sim 75$\AA~($z \sim 1.5$). The line flux limit does affect the deep survey such that low-EW systems will also be missed with increasing redshift at fixed stellar mass. However, we do predict that $>10^{11}$ \msol~systems will still be detectable at the minimum observed EW$\sim 22$\AA~at the highest redshifts probed.

For a $\sim 2000$ deg$^2$ and a line flux limit of $10^{-16}$ \cgsline~comparable to the High Latitude Spectroscopic Survey (HLSS; \citealt{Wang2022}), we predict a total of $\sim 17.6$ million \ha$+$\nii~emitters at $0.5 < z < 1.9$ and $\sim 7.5$ million \ha$+$\nii~emitters at $1 < z < 1.9$ would be observed. This is a factor 1.6 lower than the predictions of \citet{Wang2022} which predict $\sim 12$ million $1 < z < 2$ \ha$+$\nii~emitters at $10^{-16}$ \cgsline~(6.5$\sigma$) and is based on semi-analytical modeling (Galacticus; \citealt{Benson2012}) coupled with an $N$-body simulation (UNIT; \citealt{Chuang2019}) and calibrated to best match the number counts from WISP \citep{Mehta2015} and HiZELS \citep{Sobral2013}. If we remove the limiting EW and inclusion of the grism throughput, then our predictions are $\sim 12$ million $1 < z < 1.9$ \ha$+$\nii~emitters fully consistent with \citet{Wang2022}. This highlights the importance of taking into account limiting resolution and observational efficiency in number count predictions. Based on the line flux limit, HLSS will potentially observe $\gtrsim 0.2 L^\star(z = 0.5)$ and $\gtrsim 1.2 L^\star(z = 1.9)$ \ha~emitters which will strongly constrain not only the bright-end of the luminosity function, but also part of the faint-end at lower redshifts given the large sample size. Besides star-forming galaxies, HLSS will also pickup potential AGNs with the fraction rising to $\sim 15$ percent at the highest redshift end \citep{Sobral2016}. Overall, our predictions are consistent in that HLSS will have a statistically significant sample of \ha$+$\nii~emitters that are primarily star-forming and would constrain key statistical properties near cosmic noon.

\subsection{Euclid}

{\it Euclid} is a 2.3-meter space observatory that launched in July 2023 and will be able to observe $\sim 0.55$ deg$^2$ in a single pointing with its two main instruments: Visual Imaging Channel (VIS) and Near-Infrared Spectrometer and Photometer (NISP). The latter includes a slitless spectroscopy mode (NISP-S) with two main grism configurations: a `blue' grism with a single grating (BGS) and three `red' grisms (RGS) that cover the same wavelength range but configured to have three different dispersion directions (0$^\circ$, 180$^\circ$, and 270$^\circ$) to decrease contamination from overlapping spectra. The `blue' grism covers $0.92 - 1.25\micron$ with an effective wavelength of $\sim 1.1\micron$ and the `red' grisms will cover $1.25 - 1.85\micron$ with an effective wavelength of $\sim 1.5 \micron$. This corresponds to \ha~emitters at $0.4 < z < 0.9$ and $0.9 < z < 1.8$ for the blue and red grisms, respectively. Both grisms will have resolution of $R \sim 450$ for a $0.5''$ diameter source corresponding to an observed EW threshold of $\sim 27$\AA. 

Figure \ref{fig:zdistrib_surveys} shows the predicted redshift distribution using the Blue and Red grism and tabulatated in Tables \ref{table:zdistrib_Euclid_Blue} and \ref{table:zdistrib_Euclid_Red}, respectively. We predict that {\it Euclid}/Blue will be able to observe a total of $\sim 2250$, $\sim 6000$, and $\sim 12000$ $0.4 < z < 1$ \ha~emitters per deg$^2$ in a bright, intermediate, and deep survey, respectively, and {\it Euclid}/Red will be able to observe $\sim 1350$, $\sim 6200$, and $\sim 18500$ $0.9 < z < 1.9$ \ha$+$\nii~emitters per deg$^2$. Figure \ref{fig:EWdistrib_surveys} shows that both {\it Euclid}/Blue and Red grism surveys can cover a wider stellar mass and EW range with the deep pointing capturing more high EW, low-mass systems compared to the intermediate pointing. Especially towards high-$z$ in the Red grism, we predict in a 10 deg$^2$ field {\it Euclid} would be able to capture numerous $>1000$\AA~\ha~emitters for dwarf-like systems ($<10^8$ \msol) while also capturing massive star-forming galaxies ($>10^{10}$ \msol) with rest-frame \ha$+$\nii~EW up to $\sim 400$\AA.

The {\it Euclid} Wide Survey \citep{Scaramella2022} is one of the main planned surveys which will include NISP Red grism observations over $\sim 14500$ deg$^2$ area within a six year time frame and down to an observed flux limit of $2\times10^{-16}$ \cgsline. Based on our models, we predict that  {\it Euclid} Wide will be able to observe $\sim 1350$ \ha~emitters per deg$^2$ resulting in a sample size of 19.5 million $0.9 < z < 1.9 $ \ha~emitters. Given the bright emission line flux limit, {\it Euclid} Wide will be sensitive to $\sim L^\star$  and $\sim 2.5 L^\star$ \ha~emitters at $z \sim 1$ and $\sim 2$, respectively. However, the large area coverage will provide one of the strongest constraints on the bright-end of the \ha~luminosity function and will potentially have $\sim 15  - 25$ percent of the sample as AGN \citep{Sobral2016}.

\section{Discussion}
\label{sec:discussion}

\subsection{Implications for Survey Planning}

Our \wom~model in conjunction with a set of observing parameters (e.g., grism throughput, redshift range, limiting EW, and line flux threshold) can provide an assessment of estimated number counts and redshift distributions. Our approach also provides information regarding the potential range of stellar masses, \ha~line fluxes/luminosities, and \ha~EW that could be observed in a given assumed survey. This extra amount of detail not only provides accurate number count predictions by taking into account factors such as a limiting EW threshold in grism surveys, but also gauges the range of star-forming galaxy properties that could be potentially observed in a given planned survey.

 For example, {\it Roman} and {\it Euclid} show great promise in selecting a wide range of \ha~emitters within a 10 deg$^2$ survey as shown in Figure \ref{fig:EWdistrib_surveys} reaching down to 10$^{7.5}$ \msol~galaxies at $z \sim 1.5$ with rest-frame \ha~EW $>1000$\AA. The planned wide field surveys with both observatories will be able to capture orders-of-magnitude more low-mass dwarf galaxy systems near cosmic noon especially numerous cases of high EW systems which may signify bursty, episodic star-formation histories. Such systems could be analogs of potentially ionizing sources in the $z > 6$ Universe (e.g., \citealt{Stark2015,Matthee2017b,Emami2020,Atek2022}).

Another science case that can be planned using our models is the search for high mass, high EW systems which are rare and require large areas to probe. For example, Figure \ref{fig:EWdistrib_surveys} shows that at 10 deg$^2$ we predict a survey would observe many 10$^{11}$ \msol~galaxies exhibiting \ha~EW $> 100$\AA. Although not as extreme to low-mass galaxies, but in respect to the typical \ha~EW at $10^{11}$ \msol~and the corresponding redshift, these sources are indeed `extreme'. However, such sources may not appear within a $\sim 1$ deg$^2$ where the comoving volume probed is significantly reduced. Using our models, one could gauge what survey areas are needed to ensure that such sources could potentially be observed.

The two science cases defined above are examples of how our models can be used to aid in survey planning and defining science cases. However, we do caution readers that our predictions assume that the \ha~EW distribution is best described by an exponential distribution which, as we also found in \citet{Khostovan2021}, is not entirely true at extreme EW ($>1000$s EW) where outliers from the exponential distribution are found. These outliers will also have uncertain EWs given the poorly constrained faint continuum and contribute a small fraction (e.g., $<1$\% for the $z = 0.47$ LAGER sample; \citealt{Khostovan2021}). Overall, our \wom~model makes a great addition to survey planning, presents accurate number count predictions by taking into account the limiting EW threshold in grism surveys, and includes added information regarding the range of galaxy properties that may be observed.

\subsection{The Major Role of Starburst Galaxies in cosmic star formation history}

In \S\ref{sec:results_SFRD}, we show that $z > 1.5$ \ha~emitters with EW$_0 > 200$\AA~contribute $\sim 40$ percent of the overall cosmic star-formation activity and \ha~emitters with sSFR $\sim 3 - 100$ Gyr$^{-1}$ also dominate the overall cosmic star formation activity at $z > 1.5$. The rapid evolution of the cosmic SFRD for $>200$\AA~\ha~systems is consistent with past studies that have found such high EW systems to be rare in the local/low-$z$ Universe (e.g., \citealt{Lee2009,Rosenwasser2022}) but are numerous in the $z \gtrsim 1$ Universe (e.g., \citealt{vanderWel2011,Smit2014,Maseda2018}). It also provides supporting evidence of how bursty star-formation activity may be a dominate mode of star-formation in the high-$z$ Universe.

Past studies report varying results on the relative contribution of starburst galaxies to cosmic SFRD. \citet{Rodighiero2010} reports $\sim 10$\% contribution from $>10^{10}$ \msol~starburst galaxies selected at $1.5 < z < 2.5$ using \textit{Herscel}/PACS. \citet{Lemaux2014} used $\sim 2000$ \textit{Hershcel}/SPIRE-selected $0 < z < 4$ galaxies and finds an upper limit of $\sim 28$\% and $\sim 61$\% at $0 < z < 0.5$ and $2 < z < 4$, respectively, with a $>10^{10}$ \msol~stellar mass limit. We report $\sim 34$\% contribution from $10^{10 - 11.5}$ \msol~$z \sim 2.2$ \ha~emitters with no EW cut as shown in Figure \ref{fig:SFRD_ssfr_mass} and Table \ref{table:SFRD_stellar_mass}. At $z \sim 2$, the equivalent width distributions range between \wo$\sim 30 - 70$\AA~for $10^{10 - 11.5}$ \msol~galaxies with $>100$\AA~($>200$\AA)~corresponding to $\sim 20$\% ($5$\%) of the total $>10^{10}$ \msol~\ha~population and a cosmic SFRD contribution of $\sim 7$\% ($2$\%) in respect to all star-forming galaxies. The higher fractions observed in {\it Herschel} studies could be primarily due to selection of dusty star-forming galaxies which are either missed or have their dust corrections underestimated in narrowband-selected \ha~surveys. This may also be due to a volume-effect where high EW, massive galaxies are missed due to their rarity (e.g., `Ando effect'; \citealt{Ando2006}). Overall, the main consensus at $z < 2$ based on past studies and the results in this paper is that massive, starburst galaxies play a minor role in overall cosmic star formation activity.

The vast majority of $z > 1.5$ \ha~emitters with high rest-frame EW are found within the low-mass regime which is expected given the EW$_0$ -- Stellar Mass anticorrelation (see \S\ref{sec:EW_stellar_mass}). \citet{Atek2014} used {\it HST} grism spectroscopy of $\sim 1000$ \ha~emitters at $0.3 < z < 2$ and report a SFRD contribution of $13$\%, $18$\%, and $34$\% for $EW_0 > 300$, $200$, and $100$\AA, respectively, for $M \sim 10^{8.2 - 10}$ \msol~and SFR $> 2$ ($10$) \msol~yr$^{-1}$ at $z \sim 1$ ($\sim 2$). We find $> 100$\AA~and $> 200$\AA~\ha~emitters contribute 65.7\% and 39.4\%, respectively, at $z \sim 2$; however, limiting this to the same range as \citet{Atek2014} results in $\sim 31.2$\% and $18.6$\%, respectively, which is in strong agreement with \citet{Atek2014}. This also highlights how the majority of SFRD contribution at high EW is coming from $<10^{10}$ \msol~emitters where the contribution at $z \sim 2.2$ is $\sim 43$\% and  $26$\% for EW$_0 > 100$\AA~and $200$\AA, respectively, and dominated by low-mass dwarf galaxies which will have higher number densities and \ha~equivalent widths in comparison to high mass systems.

In the $z > 2$ Universe, studies suggest an even higher contribution. \citet{Caputi2017} find $>50$\% contribution of starbursts to cosmic SFRD at $3.9 < z < 4.9$ measured using {\it Spitzer}/IRAC 3.6\micron~excess consistent with strong \ha~emission. \citet{Faisst2019} finds that the majority of $z \sim 4.5$ galaxies, especially at $<10^{10}$\msol, have undergone a major bursty phase within the last 50 Myr period with a factor $>5$ enhancement in their SFR. \citet{Cohn2018} also find similar results for $2.5 < z < 4$ EELGs suggesting a recent $\sim 50$ Myr burst SF phase. Spectroscopic observations of low mass, high EW systems also find the presence of violent bursts (e.g., \citealt{Maseda2013}) supported by high gas fractions (e.g., \citealt{Maseda2014}). \citet{Tran2020} suggests that strong EWs and a bursty phase is quite common in the high-$z$ Universe and is common evolutionary path. Galaxies undergo an initial phase of bursty star formation activity resulting in faint continuum and high EWs. The continuum brightness increases due to the build-up of stellar mass after each subsequent burst resulting in a decreasing EW with time. This could explain the decrease in \ha~EW with decreasing redshift that we also find in our study. Overall, our results suggest that starburst galaxies (interpreted as high EW/sSFR systems) are important contributors to cosmic SF activity in the high-$z$ Universe dominated by low-mass, dwarf galaxies and that such a SF mode may be the dominant path of star-formation in the $z > 2$ Universe.

\subsection{High EW H$\boldmath{\alpha}$ Emitters may Reionize Universe}

Our results show an increase in the \ha~EW up to $z \sim 2.2$ with an emphasis on low-mass, high EW systems contributing a significant portion of cosmic SF activity around cosmic noon. The elevated level of star-formation suggests that such galaxies contain significant populations of massive, luminous, short-lived stars (e.g., \textit{O} and \textit{B} type stars) which would be pumping ionizing radiation into the surrounding medium. \citet{Matthee2017b} show a correlation at $z \sim 2 $ where increasing \ha~EW corresponds to an increase in the ionizing photon production efficiency, \ionxi. This has also been observed by other studies at $z \sim 2$ \citep{Tang2019,Emami2020,Nanayakkara2020} and $z \sim 4.5$ \citep{Faisst2019,Lam2019}. Recently, \citet{Atek2022} used {\it HST} grism spectroscopy and UV imaging of $0.7 < z < 1.5$ \ha~emitters and found elevated \ha/UV ratios consistent with bursty star-formation activity (see \citealt{Rezaee2022} for caveats of \ha/UV as tracer of burstiness) primarily in $<10^{8.5}$ \msol~emitters. These sources also had increasing EW with increasing \ha/UV and decreasing stellar mass. \citet{Atek2022} also reported a correlation between \ionxi~and \ha~EW such that low-mass, faint \ha~emitting star-forming galaxies may be probable sources that reionized the Universe at the Epoch of Reionization. Our results would then suggest that the typical \ionxi~of star-forming galaxies would be high, especially for low-mass, high \ha~EW starburst galaxies. Given that low-mass galaxies are more abundant than high-mass galaxies and the increasing EW with redshift at a given stellar mass, it is then probable that such galaxies are likely sources that play an important role in reionizing the Universe.

\subsection{Implications for low-$z$ interlopers in Ly$\alpha$ Samples at $z > 6$}

Investigating how galaxies formed and evolved to the present-day structures we see in the local Universe requires us to understand the high-$z$ Universe, especially at the Epoch of Reionization, where we expect the first generation of galaxies to have formed. There are several approaches to probe this cosmic regime with one popular method being the selection of \lya~emitting galaxies via narrowband-surveys such as LALA \citep{Rhoads2000,Malhotra2002,Wang2009}, SXDS \citep{Ouchi2008,Ota2010}, HiZELS \citep{Sobral2009b,Matthee2015}, CF-HiZELS \citep{Matthee2014}, CALYMHA \citep{Sobral2017}, SILVERRUSH \citep{Konno2018,Shibuya2018}, SC4K \citep{Sobral2018,Khostovan2019}, J-PLUS \citep{Spinoso2020}, LAGER \citep{Hu2019,Wold2022}, and other surveys (e.g., \citealt{Cowie1998,Kudritzki2000,Kashikawa2006,Shimasaku2006,Gronwall2007,Ota2008,Nakajima2012,Konno2016,Santos2016,Ota2017,Cabello2022}). The technique relies on both deep narrowband and broadband photometry to differentiate between high-$z$ \lya~emitters and low-$z$ foreground ELGs, such as \ha~emitters and include a minimum EW threshold. However, low-$z$ foreground contaminants are always an issue plaguing not only narrowband-selected \lya~samples, but \lya-selected samples in general. 

One efficient way to gauge the level of contamination that would arise in narrowband-selected \lya~samples is by assessing the number of high EW foreground emitters that could mimic low EW high-$z$ \lya~emitters. For example, let's assume two different $z \sim 6.6$ narrowband \lya~surveys with the condition that Survey A has \ewr(\lya)$ > 30$\AA~and Survey B has \ewr(\lya)$ > 10$\AA. Now let's assume that there maybe \ha~foreground emitters misidentified as \lya~emitters. The corresponding \ewr(\ha+\nii) of these contaminants would be $163$\AA~and $54$\AA~for Survey A and B, respectively. Assuming that the median stellar mass of the \ha~contaminants is $\sim 10^8$ \msol~(e.g., \citealt{Sobral2014}) and using Equation \ref{eqn:EW_model} with the best-fit \ha+\nii~parameters in Table \ref{table:best_model}, then the probability of having \ha~emitters with \ewr$>163$\AA~and $>54$\AA~is $\sim 1.2$\% and $\sim 23$\%. This implies that a low EW threshold survey, such as Survey B in our example, would be able to capture a larger number of \lya~emitters given the lower EW threshold, but would also significantly increase the risk of foreground \ha~contamination.

We can also assess the contamination of higher redshift \lya~emitters using our \wom~model. For example, let's assume a $z \sim 9$ \lya~narrowband survey with the condition that Survey A has  \ewr(\lya)$ > 50$\AA~and Survey B has \ewr(\lya)$ > 25$\AA. This would correspond to an \ewr(\ha+\nii) threshold of $240$\AA~and $121$\AA, respectively. Assuming a median stellar mass of $\sim 10^{9.5}$ \msol~(e.g., \citealt{Sobral2014}), this would correspond to a probability of having an \ha~emitter above these thresholds as $0.2$\% and $4.1$\%, respectively. However, depending on the survey design, if the median stellar mass is $\sim 10^{8.5}$ \msol, then the probabilities are $1.2$\% and $10.7$\%, respectively. 

We do note two main caveats with this approach. The first is that our \wom~model assumes that \ha~EW distributions are best-defined by an exponential distribution; however, there will be high EW outliers although the fraction is not significantly high (e.g., \citealt{Khostovan2021}) and is also not well-constrained given the faint broadband photometry used to constrain the continuum flux density. The second caveat is that this approach measures the possible foreground contamination only from \ha~emitters. There will also be \oiii~and \oii~foreground contaminants in \lya~samples. Overall, we have shown our \ha~\wom~model can be used to assess the level of \ha~foreground contamination in \lya-selected samples based on the assumed \ewr(\lya) threshold and a range of possible stellar masses of the contaminants.

\section{Conclusions}
\label{sec:conclusion}

In this paper, we have measured and constrained the intrinsic \ha~EW distributions from $z \sim 0.4$ to $2.2$, modeled them using our best-fit \wom~model, and explored the implications of our results in terms of the overall star-formation properties of \ha~emitters and its implications for future space-based grism surveys. Our main results are:
\begin{enumerate}
	\item We use the forward-modeling approach that was primarily developed in our first paper \citep{Khostovan2021} but now expanded to the HiZELS and DAWN samples where we measure the intrinsic \ha~EW distributions from $z \sim 0.4$ to $2.2$  in a uniform manner that takes into account selection and filter profile effects.
	\item An intrinsic EW -- $L_R$  and EW -- stellar mass anti-correlation is found at all redshift slices where low-mass/faint continuum \ha~emitters are found to have higher typical \ha~EWs compared to high-mass/bright continuum emitters. We find the slopes become steeper with increasing redshift signifying how low-mass galaxies experience a stronger, rapid evolution in their typical \ha~EWs compared to high-mass galaxies.
	\item We develop a combined double power-law model, \wom, that is used to constrain both the typical \ha~EW redshift evolution and its anti-correlation with stellar mass. We find that for a $10^{10}$ \msol~\ha~emitter, the typical \ewr~at $z = 0$ is $9.26^{+1.71}_{-1.53}$ \AA~and increases as $(1+z)^{1.78^{+0.22}_{-0.23}}$ such that by $z \sim 2.2$ the typical \ewr~is $\sim 73$\AA. The stellar mass anti-correlation slope at $z = 0$ is $-0.10^{+0.07}_{-0.07}$ and becomes steeper with redshift as $-0.05(1+z)$ such that by $z \sim 2.2$ the anti-correlation slope is $-0.26$ signifying the importance of low-mass emitters and their \ha~EW evolution.
	\item Our intrinsic \wom~model shows typical \ewr~lower than past observations but we find the main reason is due to selection effects resulting in elevated typical EWs in past measurements. We demonstrated how applying a combination of \ewr~and line flux limits can result in \ewr~redshift evolutions consistent with past observations. We also demonstrate a similar effect in the cosmic sSFR evolution where we use our intrinsic \wom~model and find our measurements are more consistent with simulations. However, incorporating selection effects, we find elevated sSFR that is consistent with observations. This suggests that the tension between observations and simulations may be rooted from selection effects.
	\item Our best-fit \wom~model predicts \ha~luminosity functions and star-formation rate functions that are consistent with past narrowband and slitless grism surveys highlighting our model can describe both the \ha~equivalent width distributions and luminosity functions coupled with an assumed stellar mass function.
	\item Given that we can predict accurate LFs and SFRFs, we extended our work to investigate the relative contribution of \ha~emitters for varying EWs, stellar masses, and sSFRs. We find that in the $z > 1.5$ Universe, cosmic SF activity is primarily arising from galaxies with \ha~\ewr$>100$\AA~and sSFRs$>10^{-8.5}$ yr$^{-1}$ contributing $50 - 60$ percent of the total cosmic SFRD. Contribution from such sources rapidly decreases with decreasing redshift with the $0.5 < z < 1.5$ Universe primarily dominated by $25 < \textrm{\ewr} < 100$\AA~and $10^{-9.5} < \textrm{sSFR} < 10^{-8.5}$ yr$^{-1}$ systems and by $z < 0.5$ it is dominated by $< 25$\AA~and $10^{-9.5}$ yr$^{-1}$ systems. Overall, we find that low-mass galaxies with high EW and sSFR are important contributors to cosmic star-formation activity at $z > 1.5$ and most likely extend as such into the high-$z$ Universe. Such sources can potentially be analogs of high-$z$ ionizing sources responsible for the reionization of the Universe.
	\item We present number count slitless grism survey predictions for {\it Roman} and {\it Euclid} incorporating the grism throughput and limiting resolution, where the latter acts as a limiting EW~threshold. We find that both {\it Roman} and {\it Euclid} will be able to find numerous \ha~emitters over a wide range of stellar masses and EWs. With the current survey parameters of {\it Roman}/HLSS, we can expect to observe $\sim 17.6$ million $0.5 < z < 1.9$ \ha$+$\nii~emitters. The current {\it Euclid} Wide Survey will also $\sim 19.5$ million $0.9 < z < 1.9$ \ha~emitters with the main difference being the {\it Euclid} Wide Survey will be limited to $>L^\star$ \ha~emitters essentially constraining the bright-end while {\it Roman}/HLSS will reach as low as $0.2 L^\star$ at the lowest redshift. 
\end{enumerate}

Overall, our \wom~model has demonstrated two major points. First, it suggests past studies may be overestimating the typical \ha~EWs and sSFRs simply due to selection biases and other observational effects not taken into account that also cause disagreement with simulations. Second, corroborating previous results, our model suggests that low mass, high EW \ha~systems are indeed important contributors to overall star-formation activity around cosmic noon and can be important analogs of sources that reionized the Universe at $z > 6$ given their high EWs. Both {\it Roman} and {\it Euclid} will be able to observe these sources in large numbers such that the next few years will be an exciting time in studying bursty star-formation activity in unprecedented detail.

\section*{Acknowledgments}
AAK acknowledges this work was supported by an appointment to the NASA Postdoctoral Program at the Goddard Space Flight Center, administered by the Universities Space Research Association (USRA) contracted through NASA. 

\section*{Data Availability}
Data from the public HiZELS narrowband survey catalogs are available on \href{https://cdsarc.cds.unistra.fr/viz-bin/cat/J/MNRAS/428/1128}{VizieR}. Raw data for the DAWN survey are also publicly accessible via the \href{https://astroarchive.noirlab.edu/portal/search/}{NOIRLab Astro Data Archive} by searching for programs 2013A-0287, 2013B-0528, and 2013B-0236.

\appendix

\section{Survey Prediction Number Counts}

Below are the number counts based on our \wom~model discussed in \S\ref{sec:forecast} and shown in Figure \ref{fig:zdistrib_surveys}. 

\begin{table}
	\centering
	\caption{Predicted \textit{Roman}/G150 \ha+\nii~Redshift Distributions defined as $dN/dz$ per deg$^{-2}$ assuming our \wo-stellar mass model, different observed \ha+\nii~line flux selection limits, and a limiting observed EW threshold of $\sim 22$\AA~corresponding to $R \sim 461$ at 1$\mu$m. Errors on the number counts at each redshift slice are based on the combination of model parameter and Poisson errors.} 
	\label{table:zdistrib_NGRST}	
	{\renewcommand{\arraystretch}{1.3}
	\begin{tabular*}{\columnwidth}{@{\extracolsep{\fill}} c c c c c}
		\hline
		& & \multicolumn{3}{c}{$F_L > F_{lim}$ (erg s$^{-1}$ cm$^{-2}$)} \\
		$z_{min}$ & $z_{max}$ & $(> 5\times10^{-17})$ & $(> 1\times10^{-16})$  & $(> 2\times10^{-16})$ \\
		\hline
		0.4 & 0.5 & $2290^{+152}_{-318}$ & $1100^{+105}_{-226}$ & $430^{+66}_{-146}$ \\
		0.5 & 0.6 & $13230^{+635}_{-410}$ & $6280^{+524}_{-254}$ & $2170^{+408}_{-163}$ \\
		0.6 & 0.7 & $21170^{+484}_{-880}$ & $9810^{+548}_{-361}$ & $3250^{+494}_{-201}$ \\
		0.7 & 0.8 & $25670^{+880}_{-584}$ & $11760^{+624}_{-378}$ & $3910^{+385}_{-210}$ \\
		0.8 & 0.9 & $26590^{+1596}_{-704}$ & $11600^{+888}_{-368}$ & $3600^{+434}_{-190}$ \\
		0.9 & 1.0 & $26930^{+674}_{-786}$ & $11000^{+512}_{-428}$ & $3080^{+322}_{-193}$ \\
		1.0 & 1.1 & $24170^{+888}_{-547}$ & $9110^{+610}_{-312}$ & $2480^{+169}_{-287}$ \\
		1.1 & 1.2 & $21340^{+611}_{-688}$ & $7300^{+525}_{-282}$ & $1670^{+183}_{-183}$ \\
		1.2 & 1.3 & $17940^{+767}_{-468}$ & $5650^{+483}_{-245}$ & $1110^{+160}_{-145}$ \\
		1.3 & 1.4 & $15980^{+408}_{-948}$ & $4700^{+225}_{-429}$ & $830^{+92}_{-193}$ \\
		1.4 & 1.5 & $13450^{+400}_{-678}$ & $3830^{+220}_{-518}$ & $460^{+129}_{-74}$ \\
		1.5 & 1.6 & $11630^{+373}_{-622}$ & $2630^{+341}_{-165}$ & $280^{+122}_{-54}$ \\
		1.6 & 1.7 & $9280^{+1045}_{-479}$ & $1810^{+527}_{-249}$ & $130^{+145}_{-54}$ \\
		1.7 & 1.8 & $7720^{+772}_{-321}$ & $1590^{+155}_{-204}$ & $170^{+42}_{-90}$ \\
		1.8 & 1.9 & $5870^{+378}_{-308}$ & $1040^{+102}_{-224}$ & $80^{+28}_{-57}$ \\
		\hline
		0.5 & 1.0 & $22718^{+1328}_{-979}$ & $10090^{+897}_{-512}$ & $3202^{+583}_{-271}$ \\
		1.0 & 1.9 & $14153^{+706}_{-625}$ & $4184^{+416}_{-327}$ & $801^{+136}_{-155}$ \\
		0.5 & 1.9 & $17212^{+656}_{-532}$ & $6293^{+417}_{-279}$ & $1658^{+225}_{-139}$ \\
		\hline
	\end{tabular*}
	}
\end{table}

\begin{table}
	\centering
	\caption{Predicted {\it Euclid}/NISP Red \ha+\nii~Redshift Distributions defined as $dN/dz$ per deg$^{-2}$ assuming our \wo-stellar mass model. Our predictions incorporate a limiting observed EW threshold of $\sim 27$\AA ~corresponding to $R \sim 450$. Errors on the number counts at each redshift slice are based on the combination of model parameter and Poisson errors.} 
	\label{table:zdistrib_Euclid_Red}
	{\renewcommand{\arraystretch}{1.3}
	\begin{tabular*}{\columnwidth}{@{\extracolsep{\fill}} c c c c c}
		\hline
		& & \multicolumn{3}{c}{$F_L > F_{lim}$ (erg s$^{-1}$ cm$^{-2}$)} \\
		$z_{min}$ & $z_{max}$ & $(> 5\times10^{-17})$ & $(> 1\times10^{-16})$  & $(> 2\times10^{-16})$ \\
		\hline
		0.8 & 0.9 & $5650^{+590}_{-245}$ & $2170^{+334}_{-148}$ & $540^{+149}_{-76}$ \\
		0.9 & 1.0 & $23790^{+896}_{-531}$ & $10060^{+317}_{-705}$ & $3050^{+281}_{-568}$ \\
		1.0 & 1.1 & $24320^{+731}_{-654}$ & $9500^{+385}_{-489}$ & $2390^{+253}_{-196}$ \\
		1.1 & 1.2 & $23770^{+531}_{-878}$ & $8890^{+301}_{-617}$ & $2000^{+245}_{-168}$ \\
		1.2 & 1.3 & $21150^{+1211}_{-510}$ & $6850^{+1024}_{-512}$ & $1470^{+324}_{-131}$ \\
		1.3 & 1.4 & $20550^{+457}_{-955}$ & $6840^{+263}_{-547}$ & $1440^{+121}_{-268}$ \\
		1.4 & 1.5 & $18140^{+802}_{-459}$ & $5480^{+498}_{-236}$ & $1100^{+107}_{-208}$ \\
		1.5 & 1.6 & $17030^{+770}_{-450}$ & $5010^{+328}_{-307}$ & $820^{+128}_{-128}$ \\
		1.6 & 1.7 & $15940^{+688}_{-471}$ & $4290^{+365}_{-235}$ & $620^{+127}_{-99}$ \\
		1.7 & 1.8 & $15260^{+393}_{-828}$ & $3870^{+225}_{-342}$ & $480^{+106}_{-92}$ \\
		1.8 & 1.9 & $5390^{+395}_{-276}$ & $980^{+373}_{-171}$ & $100^{+77}_{-32}$ \\
		\hline
		0.9 & 1.9 & $18534^{+729}_{-636}$ & $6177^{+462}_{-450}$ & $1347^{+195}_{-235}$ \\
		\hline
	\end{tabular*}
	}
\end{table}

\begin{table}
	\centering
	\caption{Predicted {\it Euclid}/NISP Blue \ha+\nii~Redshift Distributions defined as $dN/dz$ per deg$^{-2}$ assuming our \wo-stellar mass model. Our predictions incorporate a limiting observed EW threshold of $\sim 27$\AA ~corresponding to $R \sim 450$. Errors on the number counts at each redshift slice are based on the combination of model parameter and Poisson errors.} 
	\label{table:zdistrib_Euclid_Blue}
	{\renewcommand{\arraystretch}{1.3}
	\begin{tabular*}{\columnwidth}{@{\extracolsep{\fill}} c c c c c}
		\hline
		& & \multicolumn{3}{c}{$F_L > F_{lim}$ (erg s$^{-1}$ cm$^{-2}$)} \\
		$z_{min}$ & $z_{max}$ & $(> 5\times10^{-17})$ & $(> 1\times10^{-16})$  & $(> 2\times10^{-16})$ \\
		\hline
		0.4 & 0.5 & $13690^{+575}_{-470}$ & $7760^{+296}_{-521}$ & $3590^{+290}_{-242}$ \\
		0.5 & 0.6 & $22530^{+498}_{-1171}$ & $12570^{+680}_{-1309}$ & $5530^{+296}_{-672}$ \\
		0.6 & 0.7 & $24250^{+1026}_{-499}$ & $12010^{+799}_{-347}$ & $4500^{+599}_{-244}$ \\
		0.7 & 0.8 & $26540^{+746}_{-697}$ & $13090^{+404}_{-942}$ & $4580^{+220}_{-419}$ \\
		0.8 & 0.9 & $26820^{+1004}_{-542}$ & $11810^{+683}_{-355}$ & $3880^{+248}_{-318}$ \\
		0.9 & 1.0 & $6050^{+575}_{-247}$ & $2050^{+314}_{-145}$ & $420^{+136}_{-68}$ \\
		\hline
		0.4 & 1.0 & $19980^{+989}_{-862}$ & $9881^{+729}_{-933}$ & $3750^{+427}_{-486}$ \\
		\hline
	\end{tabular*}
	}
\end{table}

\bibliography{HA_EW_evolution}
\bsp	% typesetting comment
\label{lastpage}
\end{document}